\documentclass[aps,twocolumn ,nofootinbib,geometry,preprintnumbers,prd]{revtex4-1}

\usepackage{amsmath,amssymb}
\usepackage{accents}
\usepackage{nccmath}
\usepackage{color}
\usepackage[bookmarks=true,hyperfigures=true]{hyperref}
\usepackage{graphicx}
\usepackage[export]{adjustbox}
\usepackage{tensor}
\usepackage{empheq}
\usepackage{nicefrac}
\usepackage{shuffle}
\usepackage{slashed}
\usepackage[caption=false, justification=raggedright]{subfig}
\usepackage{bbold}

\usepackage[com pat=1.1.0]{tikz-feynman}   

\usetikzlibrary{snakes}

\tikzfeynmanset{
    sdot/.style={
        /tikzfeynman/dot,
        /tikz/minimum size=3pt,
    },
    graviton/.style={decorate, double, double distance=.8pt,
        decoration={snake,amplitude=2pt, segment length=8pt}
    },
    wavefunc/.style={
        /tikzfeynman/fermion, arrow size=1.1pt, color=blue},
    compact arrow/.style={
        /tikzfeynman/momentum/.cd, arrow distance=5pt, label distance=-6pt, arrow shorten=0.25
    },
	small blob/.style={
        /tikzfeynman/blob,
        /tikz/minimum size=14pt,
    }
}

\allowdisplaybreaks[2]

\newcommand{\dd}{\mathrm{d}}
\newcommand{\ccdot}{\! \cdot \!}
\def\ddelta{\delta\!\!\!{}^-\!}
\newcommand{\eqn}[1]{(\ref{#1})}

\begin{document}
\thispagestyle{empty}

\preprint{
    HU-EP-21/31,
    SAGEX-21-20-E}

\title{Classical Double Copy of Worldline Quantum Field Theory}

\author{Canxin Shi}
\email{canxin.shi@physik.hu-berlin.de}

\author{Jan Plefka}
\email{jan.plefka@hu-berlin.de}

\affiliation{Institut f\"ur Physik und IRIS Adlershof, Humboldt-Universit\"at zu Berlin,
    Zum Gro{\ss}en Windkanal 2, 12489 Berlin, Germany}

\begin{abstract}

The recently developed worldline quantum field theory (WQFT) formalism for the classical gravitational
scattering of massive bodies is extended to
massive, charged point particles coupling to bi-adjoint scalar field theory, Yang-Mills theory, and dilaton-gravity.
We establish a classical double copy relation in these WQFTs for classical observables (deflection, radiation).
The bi-adjoint scalar field theory fixes the locality structure of the double copy from Yang-Mills to dilaton-gravity.
Using this the eikonal scattering phase (or free energy of the WQFT) is computed to next-to-leading order (NLO) in coupling constants using the double copy as well as directly finding full agreement. We clarify the relation of our approach to previous studies in the effective field theory formalism.
Finally, the equivalence of the WQFT double copy to the double copy relation of the classical limit of
quantum scattering amplitudes is shown explicitly up to NLO.
\end{abstract}

\maketitle

\section{Introduction}

Finding a unified field theory of gauge bosons and gravitons has been a holy grail of physics since the days of Einstein, Heisenberg and Pauli.
While Yang-Mills theory and Einstein gravity share common features such as local symmetries, their actions
appear to be very different. In particular, their perturbative quantization in weakly interacting situations
strongly differ: while the Feynman diagrammatic expansion for gravity is notoriously involved and not renormalizable,
the Yang-Mills case is under good control and the basis of high precision predictions for scattering experiments at ever increasing orders in perturbation theory.
This is why the surprising construction of Bern-Carrasco-Johansson \cite{Bern:2008qj,Bern:2010ue} for the integrand of quantum gravitational
scattering amplitudes in terms of a double copy of Yang-Mills ones has been highly inspirational and points to a surprisingly
direct and intimate connection between these two fundamental theories of nature, see \cite{Bern:2019prr} for a recent review.
Concretely, the double copy construction arranges the building blocks of gluon scattering amplitudes in terms of kinematic numerators,
color factors and scalar propagators in such a fashion, that the kinematic numerators obey identities akin to
the Jacobi identity constitutional for the color factors resulting from the color gauge symmetry. Replacing the color factors by the
thereby identified kinematic numerators of the gluon amplitudes then
yields the integrands of scattering amplitudes in axion-dilaton-gravity (or $\mathcal{N}=0$ supergravity, i.e.~the low energy limit
of the bosonic string). Formal proofs of the double copy have been provided at tree-level
 using a variety of methods.
This ``color-kinematic duality'' has been extended to a large class of theories and the question of which (gravitational) theories
admit a double copy is an interesting and in general open one.
At the same time the double copy presents a very efficient tool to construct amplitude integrands in (super)gravity to very high loop orders, see \cite{Bern:2019prr} for an account. Yet, its deeper nature, in particular the nature of the `kinematic algebra', remains ill understood and is a subject of intense research in the modern amplitude program, see \cite{Dixon:1996wi,*Elvang:2013cua,*Henn:2014yza} for reviews.

The double copy relation for quantum scattering amplitudes leads to a natural challenge for classical general relativity. Namely, is there a classical double
copy transforming solutions of Yang-Mills (or Maxwell's) theory to gravity (with a dilaton) as well? In fact a number of such constructions
has been provided \cite{Monteiro:2014cda,*Luna:2015paa,*Luna:2016hge,*Carrillo-Gonzalez:2017iyj,*Guevara:2020xjx,*Monteiro:2020plf,*White:2020sfn,*Chacon:2021wbr,*Godazgar:2021iae}, most prominently perhaps for the Kerr-Newman solution of a spinning black-hole \cite{Arkani-Hamed:2019ymq}.
Yet, from the quantum origin of the color-kinematic
duality it should be clear that the \emph{perturbative} nature of the classical solution, i.e.~the systematic expansion about a
flat Minkowski background (known as the post-Minkowskian expansion in general relativity), should be central for the existence of a classical double copy prescription. Fascinatingly, this is also the domain  relevant for analytic gravitational wave physics, describing the inspiral (or scattering) of two massive bodies (black-holes, neutron stars or stars). The emitted gravitational radiation in that two-body process being detectable in present and future gravitational wave detectors (without the dilaton, and for the bound system in the post-Newtonian expansion, which is a combined weak field and slow motion expansion).

The approach of constructing perturbative classical solutions via the double copy was pioneered
in \cite{Goldberger:2016iau}. Here the scattering of two point particles via dilaton and graviton interactions was constructed via a double
copy of the perturbative solutions of the Yang-Mills equations coupled to point particles carrying color charge. This was further extended and clarified to the next-to-next-to leading order (NNLO) by Shen in \cite{Shen:2018ebu}, spin effects and extensions  were studied in \cite{Goldberger:2017vcg,Goldberger:2017ogt,Goldberger:2019xef}
and the analogue problem in bi-adjoint scalar field theory was considered in \cite{Goldberger:2017frp,Bastianelli:2021rbt}. All these approaches operate at the level
of the equations of motion, i.e.~one perturbatively solves both the field and particle equations of motion.
An alternative route was taken in the works \cite{Plefka:2018dpa,Plefka:2019hmz}, involving the present authors, where
a path integral based approach was taken. Here, starting from the actions describing the coupling of massive, charged particles to Yang-Mills or dilaton-gravity the force mediating fields (gluons, dilatons and graviton) were integrated out yielding
an effective action for the point particles, thereby taking the classical $\hbar\to 0$ limit.
It was shown at LO and NLO that the resulting effective action could be obtained by a suitably generalized double copy
prescription \cite{Plefka:2018dpa} taking inspiration from the amplitudes approach. Concretely, the need for a trivalent graph structure was artificially introduced via delta functions on the worldline for higher valence worldline-bulk field vertices.
Yet, this double copy prescription was shown to break down for the effective action at the NNLO \cite{Plefka:2019hmz}.
It was speculated in \cite{Plefka:2019hmz} that the reason for this breakdown lies in the attempt of double copying a gauge-variant and off-shell quantity -- the effective action -- which is at tension with the on-shell nature of the scattering amplitude double copy.

Returning to the realm of applying quantum field theory based techniques to the post-Minkowskian perturbative gravitational
scattering problem, an approach termed worldline quantum field theory (WQFT) was put forward recently  \cite{Mogull:2020sak,Jakobsen:2021zvh} that explains the relation between the two presently common approaches employing the
classical limit of scattering amplitudes, see
e.g.~\cite{Kosower:2018adc,delaCruz:2020bbn,Bern:2019crd,Damour:2017zjx,*DiVecchia:2021bdo,*Bjerrum-Bohr:2018xdl,*Cheung:2018wkq,*Bern:2019nnu,*Bjerrum-Bohr:2021din,*Herrmann:2021tct,*Brandhuber:2021eyq,*Bern:2021dqo,*Cristofoli:2021vyo}, and the PM effective field theory (EFT) approach,
see e.g.~\cite{Porto:2016pyg,*Levi:2018nxp,*Kalin:2020mvi,*Kalin:2020fhe,*Dlapa:2021npj}.
In essence it starts out from a \emph{first quantized} description of the matter field's (scalar, spinor, vector) propagator
in a gravitational background \`a la Feynman-Schwinger and demonstrates how this leads in a classical limit to the
effective field theory description. Yet, the WQFT formalism not only is of conceptual relevance (introducing an
approximate supersymmetry in the description of spin for black holes and neutron stars \cite{Jakobsen:2021zvh}),
it also provides a very efficient tool to quickly arrive at the classical observables of the scattering process without
the need to go through an iterative solution of the equations of motion in the standard EFT approach or to deal with the subtleties of the classical
limit in the amplitudes based approach. As such the deflection, the spin-kick or the explicit gravitational Bremsstrahlung waveform   as well as the eikonal being the generating function of these have been established at NLO \cite{Jakobsen:2021smu,Jakobsen:2021lvp,Jakobsen:2021zvh}.

In this work we therefore address the conceptually important question whether a double copy prescription exists for the
WQFT as well? On the face of it this is to be expected, as the WQFT may be thought of as a (partially) first order form
of the scattering amplitude problem. For this we first settle for the relevant worldline quantum field theories in section
\ref{sec2}. This includes in particular the case of the worldline coupled bi-adjoint scalar field theory which we indeed require
in order to separate the kinematic numerators from the propagator terms in the Yang-Mills case - a prerequisite for the
double copy construction. In section \ref{sec2} we establish the relevant Feynman rules for all three WQFTs, coupling to
bi-adjoint scalars, Yang-Mills and dilaton-gravity, as well as point out the relevance of the eikonal or free energy of the
WQFT as generating function for the observable of the particle's deflections. In section \ref{Se:doublecopy} we develop the
classical double copy prescription for WQFTs and construct the eikonal at NLO level. Section \ref{sec3} is devoted to the lift
of our results for the (three-particle) eikonal to the radiative waveform and in section \ref{sec4} we detail the relation
of our WQFT double copy to the established one for the quantum scattering amplitudes. After concluding we collect our conventions
in the appendix.

\section{Worldline quantum field theories}
\label{sec2}

We wish to apply the worldline quantum field theory formalism \cite{Mogull:2020sak} to massive point particles coupled to bi-adjoint scalar field theory (BS), Yang-Mills theory (YM) and dilaton-gravity (DG).
Compactly, the actions for the three theories may be written as
\begin{align}
    \label{eq:actionWQFT}
    S^{\mathrm{WQFT}} = S^{\mathrm{BS/YM/DG}} + \sum_{i} S^\mathrm{cc/pc/pm}_i,
\end{align}
where $S^{\mathrm{BS/YM/DG}}$ is the respective field theory action and
$S^\mathrm{cc/pc/pm}_i$ the respective $i$'th particle worldline action.
Note that multiple worldlines are included in eq.\eqref{eq:actionWQFT} in order to allow for interactions.
We will now focus on the worldline actions $S^\mathrm{cc/pc/pm}_i$ and delegate the rather well known field theory actions into the Appendix \ref{Ap:convention}.

The action of a massive point charge coupled to a non-abelian gauge field $A_{\mu}^{a}$ is \cite{Bastianelli:2013pta, Bastianelli:2015iba}
\begin{align} \label{eq:actionPC}
    S^\mathrm{pc} = \!-\! \int \! \dd\tau
    \left( \frac{m}{2} \left(e^{-1} \dot{x}^2 \!+\! e \right)
    \!-\! i \Psi^\dagger \dot{\Psi}
    \!-\! g \dot{x}^\mu A_\mu^a C^a\right),
\end{align}
where $e(\tau)$ is the einbein,
and the dot over a symbol denotes a derivative with respect to $\tau$.
The ``color wave function'' $\Psi_{\alpha}(\tau)$ is an auxiliary field carrying the color degrees of freedom of the particle,
the $\alpha,\beta,\ldots=1,\ldots, d_{R}$ are indices of the $d_{R}$ dimensional representation of the gauge group,
and $C^a(\tau) = \Psi^{\dagger\alpha} (T^a)_{\alpha}{}^{\beta} \Psi_{\beta}$ is the associated color charge
that determines the coupling to the gauge field $A^{a}_\mu(x)$. We shall take the generators $(T^a)_{\alpha}{}^{\beta}$ to be in the fundamental of $SU(N)$ such that $d_{R}=N$ and the adjoint indices $a,b,\ldots=1,\ldots N^{2}-1$.
This action is invariant under the reparametrization of $\tau$.
The kinetic term can be transformed into the more familiar form $- m\int \dd\tau \sqrt{\dot{x}^2}$
by solving the algebraic equations of motion for the einbein $e(\tau)$ and reinserting the solution into the action.
However, for convenience we will fix $e(\tau)=1$ such that $ \dot{x}^2=1 $ and $\tau$ is then the proper time.

Similarly, the action of a worldline minimally coupled to dilaton-gravity reads
\begin{align}
\label{eq:Spmdef}
    S^{\mathrm{pm}} =
    - \frac{m}{2} \int \dd \tau
    \left( e^{-1} e^{2\kappa \varphi} g_{\mu\nu} \dot{x}^\mu \dot{x}^\nu
    + e \right),
\end{align}
where $\varphi(x)$ is the dilaton and $e(\tau)$ is again the einbein.
The coupling constant is $\kappa = \sqrt{32 \pi G}$, where $G$ is Newton's constant.
Again, upon integrating out $e(\tau)$ we arrive at the more common form of the action
$- m \int \dd \tau e^{\kappa \varphi} \sqrt{g_{\mu\nu} \dot{x}^\mu \dot{x}^\nu }$.
We gauge fix  $e(\tau)=1$ so that $e^{2\kappa \varphi} g_{\mu\nu} \dot{x}^\mu \dot{x}^\nu = 1$.
In the weak gravitational field limit we expand
\begin{align} \label{eq:perturbG}
    g_{\mu\nu} = \eta_{\mu\nu} + \kappa\,  h_{\mu\nu},
\end{align}
where $\eta_{\mu\nu}$ is the flat-space Minkowskian metric.
Because the dilaton appears as an exponent in the action, the interaction terms become cumbersome in perturbation theory of both the worldline and field theory actions.
To simplify the calculation, we adopt the gravity gauge-fixing term and field redefinitions of $\{\varphi, h_{\mu\nu}\}$ introduced in \cite{Plefka:2018dpa} eqs.~(51)-(56).
This will decouple the worldline from $\varphi$ up to quadratic order, as well as recast the cubic
self-interaction of $h_{\mu\nu}$ into a simpler form.
The redefined field theory action can be found in \eqn{eqA:dilgravfinal} of Appendix \ref{Ap:convention}.
The worldline action in terms of the redefined fields then reads
\begin{align}
    S^{\mathrm{pm}} =
    - \frac{m}{2} \int \dd \tau
    \Big(& \dot{x}^2 + \kappa h_{\mu\nu} \dot{x}^\mu \dot{x}^\nu \nonumber \\
    &+ \frac{\kappa^2}{2} h_{\mu\rho}h_{\nu}^{\ \rho} \dot{x}^\mu \dot{x}^\nu
    \Big)
    + \mathcal{O}(\kappa^3).
\end{align}
The indices are lowered or raised by the Minkowskian metric.

Finally, let us introduce the massive point particle coupling to a bi-adjoint scalar field theory.
Here, we use the bi-adjoint scalar theory to identify the double copy kernel as introduced in \cite{Shen:2018ebu}.
A point particle interacting with a bi-adjoint scalar field (WBS) is described by \cite{Shen:2018ebu,Bastianelli:2021rbt}
\begin{multline} \label{eq:Sccaction}
    S^\mathrm{cc} =
    - \int \! \dd\tau
    \Big( \frac{m}{2} \left(e^{-1} \dot{x}^2 + e \right)
    - i \Psi^\dagger \dot{\Psi}
    - i \tilde{\Psi}^\dagger \dot{\tilde{\Psi}} \\
    -e \frac{y}{m} \phi_{a \tilde{a}} C^{a} \tilde{C}^{\tilde{a}} \Big),
\end{multline}
where $y$ is the coupling constant and $\phi_{a \tilde{a}}(x)$ is the bi-adjoint scalar field carrying two distinct color indices $a$
and $\tilde a$ related to the color and dual-color gauge groups respectively.
$\Psi_{\alpha}(\tau)$ and $\tilde{\Psi}_{\tilde\alpha}(\tau)$ are the color and dual color wave functions.
The corresponding charges are defined in a similar way as in $S^{\text{pc}}$ of \eqn{eq:actionPC}:
$C^a = \Psi^{\dagger} T^a \Psi$ and
$\tilde{C}^{\tilde{a}} = \tilde{\Psi}^{\dagger} \tilde{T}^{\tilde{a}} \tilde{\Psi}$.
Note that setting $e(\tau) = 1$ in this case will enforce the constraint $\dot{x}^2 + \frac{2y}{m^2} \phi_{a \tilde{a}} C^a \tilde{C}^{\tilde{a}} =1$.

In the worldline quantum field theory (WQFT) formalism, describing the scattering of two particles, we expand the coordinate $x^{\mu}(\tau)$ along a straight line trajectory background
\begin{align}
    x^\mu(\tau) &= b^\mu + v^\mu \tau + z^\mu(\tau),
\end{align}
with $v^2=1$ and the fluctuation $z^{\mu}$. Not that the straight line background solves the equations of motion in the field free scenario(s) $\phi_{a\tilde{a}}=A_{\mu}^{a}=h_{\mu\nu}=\varphi=0$.
 We take $b \cdot v = 0$ which may always be achieved upon shifting $\tau$.
As explained in \cite{Mogull:2020sak}, the physical meanings of $b^\mu$ and $v^\mu$ depend on the type of the worldline propagator
(advanced/retarded/time symmetric).
As in this work our main concern for the double copy construction is the integrand the $i\epsilon$ description of the propagators is of no direct concern.
Likewise, we decompose the color wave function in the background
\begin{align}
    \Psi(\tau) &= \psi + \varPsi(\tau),
\end{align}
where $\psi = \Psi(-\infty)=\text{const}$ is the initial condition, and $\varPsi(\tau)$ is the fluctuation that will be quantized.
Consequently, the color charge is
\begin{align}
    C^a = c^a + \psi^\dagger T^a \varPsi + \varPsi^\dagger T^a \psi + \varPsi^\dagger T^a \varPsi,
\end{align}
where we have defined the background color charge
$c^a = \psi^\dagger T^a \psi$.
A similar decomposition applies to the dual color wave function $\tilde{\Psi}(\tau)$, and all respective dual quantities are denoted with a tilde.

In WQFT, the physical observables are computed as the expectation values of the corresponding operators.
We will integrate out the BS/YM/DG fields  $\phi_{a\tilde{a}}; A_{\mu}^{a}; h_{\mu\nu},\varphi$ as well as all fluctuations of worldline degrees of freedom $z(\tau), \varPsi(\tau), \tilde{\varPsi}(\tau)$ in the path integral, so the results only depend on the background fields $b, v, \psi$.
In the path integral, the expectation value of an operator $\mathcal{O}$ is expressed as
\begin{align}\label{eq:expectation}
    \langle \mathcal{O} \rangle = \frac{1}{\mathcal{Z_\mathrm{WQFT}}} \!\int D[\Phi] \!
    \prod_i D[z_i] \left(D[\varPsi_i, \tilde{\varPsi}_i] \right)
    \mathcal{O}\, e^{i S^\mathrm{WQFT}},
\end{align}
where $\Phi \in \{ \phi_{a \tilde{a}}, A_\mu^a, h_{\mu\nu}, \varphi \}$ denotes the bosonic fields in the respective theories.
$\mathcal{Z_\mathrm{WQFT}}$ is the partition function,
\begin{align}
    \mathcal{Z}_\mathrm{WQFT} = \int D[\Phi] \!
    \prod_i D[z_i] \left(D[\varPsi_i, \tilde{\varPsi}_i] \right)
    e^{i S^\mathrm{WQFT}}.
\end{align}
In the binary case ($i=1,2$) $\mathcal{Z_\mathrm{WQFT}}$ may be identified with the exponentiated  eikonal phase $\chi$.
The momentum deflection of a particle $\Delta p_i^\mu$ can be calculated by taking the derivative of the eikonal with respect to $b_i^\mu$.
Here we claim that this relation holds for an arbitrary number of worldlines.
We will provide a simple proof in the path integral formalism.
Let us first consider the derivative of $\ln \mathcal{Z_\mathrm{WQFT}}$ with respect to $b_i^\mu$,
\begin{align} \label{eq:logZtob}
    i \frac{\partial \ln \mathcal{Z}_\mathrm{WQFT}}{\partial b_i^\mu} =
    \left\langle
     - \frac{ \partial S^\mathrm{WQFT}} {\partial b_i^\mu} \right\rangle
    =- \int_{-\infty}^{+\infty}\!\! \dd \tau
    \left\langle \frac{ \partial L^\mathrm{pp}} {\partial x_i^\mu} \right\rangle,
\end{align}
where in the last step we exploit the fact that in the full action $b_i^\mu$ only appears as the $\tau$-independent background of $x_i^\mu(\tau)$ in the point particle action  $S^{\mathrm{pp}} = \int \dd \tau L^{\mathrm{pp}}$, where $L^{\mathrm{pp}}$ is the Lagrangian.
As the expectation value of the equation of motion for $x(\tau)$ vanishes, we can rewrite eq.\eqref{eq:logZtob} as
\begin{align}
    i \frac{\partial \ln \mathcal{Z}_\mathrm{WQFT}}{\partial b_i^\mu}
    = -\int_{-\infty}^{+\infty}\!\! \dd \tau
    \left\langle  \frac{\dd }{\dd \tau} \frac{\partial L^\mathrm{pp}} {\partial \dot{x}_i^\mu} \right\rangle
    = \left. \left\langle p^{\mathrm{can}}_{i, \mu} \right\rangle \right|_{-\infty}^{+\infty},
\end{align}
where $p^{\mathrm{can}}_{i, \mu} = -\partial L^\mathrm{pp}/{\partial \dot{x}_i^\mu} $ is the canonical momentum conjugated to $x^\mu$.
Since we are studying a scattering process, in past and future infinity we may assume that the point particles are so far separated that the interaction terms vanish.
In this case $p^{\mathrm{can}}_{i, \mu}$ reduces to the kinematic momentum
$m_i \dot{x}_i^\mu$, so we have
\begin{align}
     m_i \Delta \dot{x}_i^\mu
     =
     i \frac{\partial \ln \mathcal{Z}_\mathrm{WQFT}}{\partial b_{i,\mu}}.
\end{align}
Therefore in this letter, we define the generalized eikonal phase for more than two worldlines as,
\begin{align}
    \chi = -i \ln \mathcal{Z}_\mathrm{WQFT}.
\end{align}
In section \ref{Se:doublecopy}, we will perform a double copy for the eikonal to next-to-leading order.

Since we will mostly work in momentum space, it will be useful to express the worldline fluctuations as
\begin{equation}
\begin{split}
    \label{eq:wlFourier}
    z^\mu(\tau) &= \int_{\omega} e^{-i \omega \tau} z^{\mu}(\omega), \\
    \varPsi(\tau) &= \int_{\omega} e^{-i \omega\tau} \varPsi(\omega), \\
    \varPsi^\dagger(\tau) &= \int_{\omega} e^{-i \omega \tau} \varPsi^{\dagger}(-\omega) \, .
\end{split}
\end{equation}
The dual color wave function $\tilde{\varPsi}$ in momentum space is defined in the same way as $\varPsi$.
For convenience we use the integral shorthands
$\int_\omega := \int \frac{\dd\omega}{2\pi}$,
$\int_k := \int \frac{\dd^4 k}{(2\pi)^4}$ as well as $\ddelta(\omega) := 2\pi \delta(\omega)$ and
$\ddelta^{(4)} (k^\mu) := (2\pi)^4 \delta^{(4)} (k^\mu)$.
When evaluated on the worldline, the generic field $\Phi$ may be expanded as
\begin{widetext}
    \begin{align}
        \label{eq:decompose}
        \Phi(x(\tau))
        &=\int_{k} e^{i k \cdot(b+v \tau+z(\tau))} \Phi(-k)
        = \sum_{n=0}^{\infty} \frac{i^{n}}{n !} \int_{k} e^{i k \cdot(b+v \tau)}(k \cdot z(\tau))^{n} \Phi(-k) \nonumber \\
        &=\int_{k} e^{i k \cdot b} \Phi(-k)
        \left( e^{i k \cdot v\tau}
        + i \int_\omega e^{i (k \cdot v + \omega) \tau} k \cdot z(-\omega)
        \right) + \mathcal{O}(z^2).
    \end{align}
\end{widetext}
We take the expansion only to linear order in $z^\mu$ since this is the highest term we need in this letter.
A complete expression of $h_{\mu\nu}$ to all orders in $z$ may be found in \cite{Mogull:2020sak}.

Next we extract the Feynman rules from the worldline actions.
The worldline propagators are the same in all three theories,
\begin{align}
    \label{eq:propZ}
    \begin{tikzpicture}[baseline={(0,0)}]
        \begin{feynman}
            \vertex[sdot, label=180:$z^\mu$] at (-1,0) (v1) {};
            \vertex[sdot, label=0:$z^\nu$] at (0,0) (v2) {};
            \diagram*{
                (v1) -- [ultra thick, momentum={[arrow distance=5pt]\( \omega \)}] (v2)
            };
        \end{feynman}
    \end{tikzpicture}
    &= -\frac{i}{m} \frac{\eta^{\mu \nu}}{\omega^2} \\
    \label{eq:propPsi}
    \begin{tikzpicture}[baseline={(0,0)}]
        \begin{feynman}
            \vertex[sdot, label=180:$\varPsi^\dagger$] at (-1,0) (v1) {};
            \vertex[sdot, label=0:$\varPsi$] at (0,0) (v2) {};
            \diagram*{
                (v1) -- [wavefunc, momentum={[arrow distance=5pt]\(\color{black} \omega \)}] (v2)
            };
        \end{feynman}
    \end{tikzpicture}
     &= \frac{i}{\omega}\, .
\end{align}
The propagator of the dual field $\tilde{\varPsi}$ is identical to the one for $\varPsi$.

Let us now begin with the analysis of the Yang-Mills coupled WQFT.
With \eqref{eq:wlFourier} and \eqref{eq:decompose} we can expand the interaction term of $S^{\text{pc}}$ from eq.~\eqn{eq:actionPC} as
\begin{align}
    \label{eq:WYMinteraction}
    S^{\mathrm{pc}}_{\mathrm{int}}
    =& g \int \dd \tau \, \dot{x}^\mu(\tau) \cdot A^a(x(\tau)) \, C^a(\tau) \\
    =& g \int_k e^{i k \cdot b} v\cdot A^a(-k) \ddelta(k\cdot v) c^a \nonumber\\
    &+ g \int_{k,\omega} e^{i k \cdot b} A_\mu^a(-k) \ddelta(k\cdot v + \omega) \nonumber\\ 
    &\quad\times \Big[ i \big( \omega z^\mu(-\omega) + v^\mu k\cdot z(-\omega) \big) c^a \nonumber\\
    &\quad\quad  + v^\mu (\psi^\dagger T^a \varPsi(-\omega)
    \!+\! \varPsi^\dagger(\omega) T^a \psi )  \Big]\!+\!  \mathcal{O} \left( (z, \varPsi)^2 \right)
    \nonumber
\end{align}
where we keep the interaction to linear order in worldline fluctuations.
The Feynman rules of the worldline-gluon vertices can be directly read off from \eqref{eq:WYMinteraction},
\begin{align}
    \label{eq:vertex0z1gYM}
    \begin{tikzpicture}[baseline={(current bounding box.center)}]
        \begin{feynman}
            \vertex (v1) at (-1,0) {};
            \vertex (v2) at (1,0) {};
            \vertex[sdot] at (0,0) (v0) {};
            \vertex (k) at (0,-1.3) {$A_\mu^a$};
            \diagram*{
                (v1) -- [scalar] (v0) --[scalar] (v2),
                (v0) -- [opacity=0, momentum'={\(k\)} ] (k),
                (v0) -- [gluon] (k),
            };
        \end{feynman}
    \end{tikzpicture}
    &= i g e^{i k \cdot b} \ddelta(k\cdot v) v^\mu c^a  \\
    \begin{tikzpicture}[baseline={(0,-0.8)}]
        \begin{feynman}
            \vertex at (-1,0) (v1) {};
            \vertex at (1.2,0) (v2) {$z^\rho$};
            \vertex[sdot] at (0,0) (v0) {};
            \vertex at (0,-1.3) (k) {$A_\mu^a$};
            \diagram*{
                (v1) -- [scalar] (v0) -- [ultra thick, momentum={[arrow distance=5pt]\( \omega \)}] (v2),
                (v0) -- [opacity=0, momentum'={\(k\)} ] (k),
                (v0) -- [gluon] (k),
            };
        \end{feynman}
    \end{tikzpicture}
    &
    \begin{aligned}
        = -& g e^{i k \cdot b} \ddelta(k\ccdot v + \omega) \\
        &~ \times  ( \omega \eta^{\mu\rho} + v^\mu k^\rho ) c^a
    \end{aligned} \\
    \begin{tikzpicture}[baseline={(current bounding box.center)}]
        \begin{feynman}
            \vertex at (-1,0) (v1) {};
            \vertex at (1.2,0) (v2) {$\varPsi^\dagger$};
            \vertex[sdot] at (0,0) (v0) {};
            \vertex at (0,-1.3) (k) {$A_\mu^a$};
            \diagram*{
                (v1) -- [scalar] (v0) -- [wavefunc,arrow size=1.1pt, momentum={[arrow distance=5pt]\( \color{black} \omega \)}] (v2),
                (v0) -- [opacity=0, momentum'={\(k\)} ] (k),
                (v0) -- [gluon] (k),
            };
        \end{feynman}
    \end{tikzpicture}
    &= i g e^{i k \cdot b} \ddelta(k\cdot v + \omega) v^\mu (T^a \psi) \\
    \label{eq:vertex1psiin1gYM}
    \begin{tikzpicture}[baseline={(current bounding box.center)}]
        \begin{feynman}
            \vertex at (-1,0) (v1) {$\varPsi$};
            \vertex at (1.2,0) (v2) {};
            \vertex[sdot] at (0,0) (v0) {};
            \vertex at (0,-1.3) (k) {$A_\mu^a$};
            \diagram*{
                (v1) -- [wavefunc, momentum={[arrow distance=5pt]\( \color{black} \omega \)}] (v0) -- [scalar] (v2),
                (v0) -- [opacity=0, momentum={\(k\)} ] (k),
                (v0) -- [gluon] (k),
            };
        \end{feynman}
    \end{tikzpicture}
    &= i g e^{i k \cdot b} \ddelta(k\cdot v - \omega) v^\mu (\psi^\dagger T^a).
\end{align}

Turning to the bi-adjoint scalar coupled WQFT, we can expand the worldline-scalar coupling of \eqn{eq:Sccaction} in the same way,
\begin{align}
    \label{eq:WBSinteraction}
    S^{\mathrm{cc}}_{\mathrm{int}} =& \frac{y}{m} \int \dd \tau \phi^{a \tilde{a}}(x(\tau)) C^a(\tau) C^{\tilde{a}}(\tau) \\
    =& \frac{y}{m}  \int_k e^{i k \cdot b} \phi^{a \tilde{a}}(-k)  \ddelta(k\cdot v) c^a c^{\tilde{a}} \nonumber\\
    &+ \frac{y}{m}  \int_{k,\omega} e^{i k \cdot b} \phi^{a \tilde{a}}(-k) \ddelta(k\cdot v + \omega)
    \Big[ i k\cdot z(-\omega) c^a c^{\tilde{a}}  \nonumber\\
    &\quad +  \left( \psi^\dagger T^a \varPsi(-\omega)
    + {\varPsi}^\dagger(\omega) T^{{a }} {\psi } \right) c^{\tilde{a}}\nonumber\\
    &\quad +  c^{{a}}\! \left(\tilde{\psi}^\dagger \tilde{T}^{\tilde{a}} \tilde{\varPsi}(-\omega)
    + \tilde{\varPsi}^\dagger(\omega) \tilde{T}^{\tilde{a }} \tilde{\psi } \!\right)
    \!\Big] \!+\! \mathcal{O} \left( (z, \varPsi)^2 \right) .\nonumber
\end{align}
Again, we keep only the terms that we need in this work.
From the interaction \eqref{eq:WBSinteraction} we extract the Feynman rules
\begin{align} \label{eq:vertex0z1gBS}
    \begin{tikzpicture}[baseline={(current bounding box.center)}]
        \begin{feynman}
            \vertex (v1) at (-1,0);
            \vertex (v2) at (1,0);
            \vertex[sdot] at (0,0) (v0) {};
            \vertex (k) at (0,-1.3) {$\phi^{ab}$};
            \diagram*{
                (v1) -- [scalar] (v0) --[scalar] (v2),
                (v0) -- [opacity=0, momentum'={\(k\)} ] (k),
                (v0) -- [photon] (k),
            };
        \end{feynman}
    \end{tikzpicture}
    =& 
    \frac{iy}{m} e^{i k \cdot b} \ddelta(k\cdot v) c^a c^{\tilde{a}} \\
    \begin{tikzpicture}[baseline={(current bounding box.center)}]
        \begin{feynman}
            \vertex at (-1,0) (v1) {};
            \vertex at (1.2,0) (v2) {$z^\rho$};
            \vertex[sdot] at (0,0) (v0) {};
            \vertex at (0,-1.3) (k) {$\phi^{ab}$};
            \diagram*{
                (v1) -- [scalar] (v0) -- [ultra thick, momentum={[arrow distance=5pt]\( \omega \)}] (v2),
                (v0) -- [opacity=0, momentum'={\(k\)} ] (k),
                (v0) -- [photon] (k),
            };
        \end{feynman}
    \end{tikzpicture}
    =& - \frac{y}{m}  e^{i k \cdot b} \ddelta(k\cdot v + \omega) k^\rho c^a c^{\tilde{a}}
    \end{align}
    \begin{align}
    \begin{tikzpicture}[baseline={(current bounding box.center)}]
        \begin{feynman}
            \vertex at (-1,0) (v1) {};
            \vertex at (1.2,0) (v2) {$\varPsi^\dagger$};
            \vertex[sdot] at (0,0) (v0) {};
            \vertex at (0,-1.3) (k) {$\phi^{ab}$};
            \diagram*{
                (v1) -- [scalar] (v0) -- [wavefunc,arrow size=1.1pt, momentum={[arrow distance=5pt]\( \color{black} \omega \)}] (v2),
                (v0) -- [opacity=0, momentum'={\(k\)} ] (k),
                (v0) -- [photon] (k),
            };
        \end{feynman}
    \end{tikzpicture}
    &
    = \frac{iy}{m}  e^{i k \cdot b} \ddelta(k\cdot v + \omega)
     (T^a \psi) c^{\tilde{a}}
    \label{eq:vertex1psiout1gBS}  \\
    \begin{tikzpicture}[baseline={(current bounding box.center)}]
        \begin{feynman}
            \vertex at (-1,0) (v1) {$\varPsi$};
            \vertex at (1.2,0) (v2) {};
            \vertex[sdot] at (0,0) (v0) {};
            \vertex at (0,-1.3) (k) {$\phi^{ab}$};
            \diagram*{
                (v1) -- [wavefunc, momentum={[arrow distance=5pt]\( \color{black} \omega \)}] (v0) -- [scalar] (v2),
                (v0) -- [opacity=0, momentum={\(k\)} ] (k),
                (v0) -- [photon] (k),
            };
        \end{feynman}
    \end{tikzpicture}
    &
    \!\!= \frac{i y}{m}  e^{i k \cdot b} \ddelta(k\cdot v - \omega)
    (\psi^\dagger T^a ) c^{\tilde{a}}.
    \label{eq:vertex1psiin1gBS}
\end{align}
For vertices that involves the dual wave function, we simply use \eqref{eq:vertex1psiout1gBS} or \eqref{eq:vertex1psiin1gBS} and change ${\Psi}$ to $\tilde{\Psi}$.

In the dilaton-gravity coupled WQFT, the interaction term is remarkably simplified by the field redefinitions of $\{\varphi, h_{\mu\nu} \}$.
In the end the linear order in $h_{\mu\nu}$ is no different than the interaction term of a point mass in pure gravity, which is given in \cite{Mogull:2020sak} to all orders in $z(\omega)$.
Here we provide the first terms we need in this letter,
\begin{widetext}
\begin{align}
     S^{\mathrm{pm}}_{\mathrm{int}} = &
     -\frac{m \kappa}{2} \int_{k} e^{i k \cdot b} \delta(k \cdot v) h_{\mu \nu}(-k) v^{\mu} v^{\nu}
     -i \frac{m \kappa}{2} \int_{k, \omega} e^{i k \cdot b} \delta(k \cdot v+\omega) h_{\mu \nu}(-k) z^{\rho}(-\omega)\left(2 \omega v^{(\mu} \delta_{\rho}^{\nu)}+v^{\mu} v^{\nu} k_{\rho}\right) \nonumber\\
     & -\frac{m \kappa^2}{4} \int_{k_1,k_2}
     e^{(k_1+k_2) \cdot b}\ddelta((k_1+k_2) \ccdot v)
     h_{\mu\rho}(-k_1) h_{\nu}^{\ \rho}(-k_2) v^\mu v^\nu
     + \mathcal{O}(k^3, z^2),
\end{align}
\end{widetext}
from which we obtain the Feynman rules,
\begin{align}
    \label{eq:vertex0z1hDG}
    \begin{tikzpicture}[baseline={(current bounding box.center)}]
        \begin{feynman}
            \vertex (v1) at (-1,0);
            \vertex (v2) at (1,0);
            \vertex[sdot] at (0,0) (v0) {};
            \vertex (k) at (0,-1.3) {$h_{\mu\nu}$};
            \diagram*{
                (v1) -- [scalar] (v0) --[scalar] (v2),
                (v0) -- [opacity=0, momentum'={\(k\)} ] (k),
                (v0) -- [graviton] (k),
            };
        \end{feynman}
    \end{tikzpicture}
    =& -i \frac{m \kappa}{2} e^{i k \cdot b} \ddelta(k \cdot v) v^{\mu} v^{\nu}  \\
    \begin{tikzpicture}[baseline={(0,-0.8)}]
        \begin{feynman}
            \vertex at (-1,0) (v1) {};
            \vertex at (1.2,0) (v2) {$z^\rho$};
            \vertex[sdot] at (0,0) (v0) {};
            \vertex at (0,-1.3) (k) {$h_{\mu\nu}$};
            \diagram*{
                (v1) -- [scalar] (v0) -- [ultra thick, momentum={[arrow distance=5pt]\( \omega \)}] (v2),
                (v0) -- [opacity=0, momentum'={\(k\)} ] (k),
                (v0) -- [graviton] (k),
            };
        \end{feynman}
    \end{tikzpicture}
     &
    \begin{aligned}
        =& \frac{m \kappa}{2} e^{i k \cdot b} \ddelta(k \cdot v+\omega) \\
        &  \qquad \left(2 \omega v^{(\mu} \delta_{\rho}^{\nu)}+v^{\mu} v^{\nu} k_{\rho}\right)
    \end{aligned} \\
    \label{eq:vertex2z1hDG}
    \begin{tikzpicture}[baseline={(0,-0.8)}]
        \begin{feynman}
            \vertex at (-1,0) (v1) {};
            \vertex at (1.2,0) (v2) {};
            \vertex[sdot] at (0,0) (v0) {};
            \vertex at ($(v0)!1!-40:(0,-1.4)$) (k1) {$h_{\mu\nu}$};
            \vertex at ($(v0)!0.05!-40:(0,-1.4)$) (f1) {};
            \vertex at ($(v0)!1!40:(0,-1.4)$) (k2) {$h_{\rho\sigma}$};
            \vertex at ($(v0)!0.05!40:(0,-1.4)$) (f2) {};
            \diagram*{
                (v1) -- [scalar] (v0) -- [scalar] (v2),
                (f1) -- [opacity=0, momentum'={[arrow distance=5pt, label distance=-6pt, arrow shorten=0.25] $k_1\ \ $} ] (k1),
                (v0) -- [graviton] (k1),
                (f2) -- [opacity=0, momentum={[compact arrow]$\ \ k_2$} ] (k2),
                (v0) -- [graviton] (k2),
            };
        \end{feynman}
    \end{tikzpicture}
    &
    \begin{aligned}
        = - &\frac{m \kappa^2}{2} \int_{k_1,k_2}
        e^{i(k_1+k_2) \cdot b}  \\
        &\ddelta \left((k_1 \!+\!k_2) \ccdot v\right)  v^{(\mu} \eta^{\nu)(\rho} v^{\sigma)},
    \end{aligned}
\end{align}
where the parenthesis of Lorentz indices denotes symmetrization with unit weight, e.g.~$v_1^{(\mu} v_2^{\nu)} = \frac{1}{2}(v_1^{\mu} v_2^{\nu} + v_1^{\nu} v_2^{\mu})$.

\section{Classical double copy}
\label{Se:doublecopy}
One of the main challenges of constructing the double copy in the classical limit of quantum field theories
is that the locality structure is concealed.
This is rooted in the classical limit of the massive scalar propagator \cite{Mogull:2020sak},
which contains both double and single propagators as we can see in WQFT from \eqref{eq:propZ} and \eqref{eq:propPsi}.
Following \cite{Shen:2018ebu} we tackle this difficulty by using the bi-adjoint scalar theory to identify the correct locality structure, i.e.~disentangle the kinematical numerators from the propagator terms.

Another important strategy to establish the classical double copy is to consider more than two worldlines even if we are ultimately interested only in  two-body interactions.
This is to avoid the situation where some color factors in the two-body situation are vanishing but the corresponding numerators do not, which under the double copy map may yield non-zero contributions. This may be evaded if we use as many worldlines as worldline-field interactions occur.
Specifically, we will consider an $(n+2)$-body system at $\mathrm{N^{n}LO}$.
In the WQFT formalism, this is equivalent to taking into account only tree diagrams.
To retrieve the binary system from this, we need to sum all possible ways of fusing the $(n+2)$ worldlines into $2$ worldlines.
In summary, our double copy relation of the eikonal phase reads
\begin{subequations} \label{eq:dcEikonal}
\begin{align}
    \label{eq:dcEikonalBS}
    \chi^{\mathrm{BS}}_n =& -y^{2n} \int {\dd \mu_{1,2,...,(n+1)}} \sum_{i,j} C_i K_{ij} \tilde{C}_j, \\
    \label{eq:dcEikonalYM}
    \chi^{\mathrm{YM}}_n =& -(ig)^{2n} \int {\dd  \mu_{1,2,...,(n+1)}} \sum_{i,j} C_i K_{ij} {N}_j, \\
    \label{eq:dcEikonalDG}
    \chi^{\mathrm{DG}}_n =& -\left( \frac{\kappa}{2} \right)^{2n} \int  {\dd \mu_{1,2,...,(n+1)}}\sum_{i,j} N_i K_{ij} N_j,
\end{align}
\end{subequations}
where $C_i, \tilde{C}_j$ denotes the color and dual color factors, $N_j$ are the numerators, and $K_{ij}$ are the so-called double copy kernels that encodes the locality structure.
For further convenience, we have also defined the integral measure
\begin{align}
    \dd \mu_{1,2,...,n} = \prod_{i=1}^{n} \left( \frac{\dd^4 k_i}{(2\pi)^4} e^{i k_i \cdot b_i}
    \ddelta\left(k_i \ccdot p_i \right) \right)\ddelta^{(4)} \bigg(\sum_{i=1}^n k_i^\mu \bigg),
\end{align}
where $k_i$ is the total outgoing momentum of bosonic fields $\Phi(x)$ attached to a worldline.
Note that we have defined the momentum of the massive particle as
\begin{align}
    p_i^\mu := m_i v_i^\mu,
    \quad \text{so that} \quad
    \ddelta(k_i \ccdot p_i) = \frac{\ddelta(k_i \ccdot v_i)}{m_i}.
\end{align}
Hereafter we will always express the numerator $N_j$ in terms of the momentum $p_i^\mu$ which is necessary in order to balance the mass dimension under the double copy.
The kinematic numerators $N_i$ are arranged to satisfy the same algebraic equations as the color factors $C_i$,
\begin{align}
    &C_i + C_j + C_k = 0
    \quad \Rightarrow \quad
    N_i + N_j + N_k = 0.
\end{align}
It is worth mentioning that we have the color-kinematic duality already at quartic order in the coupling constant.

\subsection{Eikonal at leading order (LO)}
The locality structure at leading order is trivial, so we do not need to employ the bi-adjoint scalar theory in order to
double copy YM color charged particles to DG ones.
In Yang-Mills coupled WQFT (WYM) the eikonal phase at this order involves only one diagram.
Using the Feynman rules \eqref{eq:vertex0z1gYM} and the gluon propagator \eqref{eq:propA}, we have
\begin{align} \label{eq:eikonalYMLO}
    i \chi^{\mathrm{YM}}_1
    =&
    \begin{tikzpicture}[baseline={(current bounding box.center)}]
    \begin{feynman}
        \vertex (v1) at (-1,0) {$1$};
        \vertex (v2) at (0.8,0);
        \vertex[sdot] at (0,0) (v0) {};
        \vertex[sdot] (w0) at (0,-1) {};
        \vertex (w1) at (-1,-1) {$2$};
        \vertex (w2) at (0.8,-1);
        \diagram*{
            (v1) -- [scalar] (v0) -- [scalar] (v2),
            (v0) -- [opacity=0, momentum'={\(k_1\)} ] (w0),
            (v0) -- [gluon] (w0),
            (w1) -- [scalar] (w0) -- [scalar] (w2)
        };
    \end{feynman}
    \end{tikzpicture}
    = ig^2 \int \dd \mu_{1,2}
    \frac{(p_1 \ccdot p_2) (c_1 \ccdot c_2) }{k_1^2}
\end{align}
where we have massaged the formula to fit the form as \eqref{eq:dcEikonalYM}.
We can identify the color factor, the numerator and the double copy kernel as
\begin{align}
    C =  (c_1 \ccdot c_2), \quad N = (p_1 \ccdot p_2), \quad K = \frac{1}{k_1^2}.
\end{align}

In worldline coupled dilaton-gravity (WDG), thanks to the decoupling of $\varphi$ from the worldline, we have also only one diagram mediated by $h_{\mu\nu}$. With \eqref{eq:vertex0z1hDG} and the graviton propagator \eqref{eq:proph}, we obtain
\begin{align}
    i\chi^{\mathrm{DG}}_1 =
    \begin{tikzpicture}[baseline={(current bounding box.center)}]
        \begin{feynman}
            \vertex (v1) at (-1,0) {$1$};
            \vertex (v2) at (0.8,0);
            \vertex[sdot] at (0,0) (v0) {};
            \vertex[sdot] (w0) at (0,-1) {};
            \vertex (w1) at (-1,-1) {$2$};
            \vertex (w2) at (0.8,-1);
            \diagram*{
                (v1) -- [scalar] (v0) -- [scalar] (v2),
                (v0) -- [opacity=0, momentum'={\(k_1\)} ] (w0),
                (v0) -- [graviton] (w0),
                (w1) -- [scalar] (w0) -- [scalar] (w2)
            };
        \end{feynman}
    \end{tikzpicture}
    = \frac{-i \kappa^2}{4} \int {\dd \mu_{1,2}}
    \frac{(p_1 \cdot p_2)^2}{k_1^2}.
\end{align}
Hence at the leading order the eikonal of Yang-Mills and dilaton gravtiy obviously possess a double copy relation \eqref{eq:dcEikonal}.

\subsection{Eikonal at Next-to-Leading order (NLO)}
As explained before, at next-to-leading order, to avoid the vanishing of some contributions in worldline coupled bi-adjoint scalar theory (WBS) and Yang-Mills coupled WQFT theory, we will consider three worldlines.
At this order the locality structure is non-trivial.
As we will see, the double copy kernel is off-diagonal.
Therefore, we will first consider the bi-adjoint scalar theory to identify the kernel.
The Feynman diagrams in WBS can be calculated using the Feynman rules \eqref{eq:vertex0z1gBS}-\eqref{eq:vertex1psiin1gBS} and the three-point vertex of $\phi_{a\tilde a}$ \eqref{eq:vertex3phi},
\begin{widetext}
\begin{align}
    \label{eq:BSNLOzprop}
    \begin{tikzpicture}[baseline={(current bounding box.center)}]
        \begin{feynman}
            \vertex [label=180:$1$] at (0,0) (a0) ;
            \vertex [sdot, right=.6cm of a0] (a1) {};
            \vertex [sdot, right=.7cm of a1] (a2) {};
            \vertex [right=.6cm of a2] (a3);
            \vertex [below=1 of a1] (i1);
            \vertex [sdot] at ($(a1)!1!-30:(i1)$) (b1) {};
            \vertex [left=.4cm of b1] (i2);
            \vertex [label=180:2] at ($(b1)!1!-30:(i2)$) (b0);
            \vertex [right=.4cm of b1] (i3);
            \vertex at ($(b1)!1!-30:(i3)$) (b2);
            \vertex [below=1 of a2] (j1);
            \vertex [sdot] at ($(a2)!1!30:(j1)$) (c1) {};
            \vertex [left=.4cm of c1] (j2);
            \vertex [label=180:3] at ($(c1)!1!30:(j2)$) (c0);
            \vertex [right=.4cm of c1] (j3);
            \vertex at ($(c1)!1!30:(j3)$) (c2);
            \diagram*{
                (a0) --[scalar] (a1) --[ultra thick] (a2) --[scalar] (a3),
                (b0) --[scalar] (b1) --[scalar] (b2),
                (c0) --[scalar] (c1) --[scalar] (c2),
                (a1) --[photon, rmomentum={[compact arrow]$\ \ k_2$} ] (b1),
                (a2) --[photon, rmomentum={[compact arrow]$\ \ k_3$} ] (c1),
            };
        \end{feynman}
    \end{tikzpicture}
    =& {-iy^4}\int \frac{\dd \mu_{1,2,3}}{k_2^2 k_3^2}
    \frac{k_2 \cdot k_3}{(k_2 \ccdot p_1)^2}
    (c_1\ccdot c_2) (c_1\ccdot c_3)
    (\tilde{c}_1\ccdot \tilde{c}_2) (\tilde{c}_1\ccdot \tilde{c}_2)
     \\
    \label{eq:BSNLOinprop}
    \begin{tikzpicture}[baseline={(current bounding box.center)}]
        \begin{feynman}
            \vertex [label=180:$1$] at (0,0) (a0);
            \vertex [sdot, right=.6cm of a0] (a1) {};
            \vertex [sdot, right=.7cm of a1] (a2) {};
            \vertex [right=.6cm of a2] (a3);
            \vertex [below=1 of a1] (i1);
            \vertex [sdot] at ($(a1)!1!-30:(i1)$) (b1) {};
            \vertex [left=.4cm of b1] (i2);
            \vertex [label=180:$2$] at ($(b1)!1!-30:(i2)$) (b0);
            \vertex [right=.4cm of b1] (i3);
            \vertex at ($(b1)!1!-30:(i3)$) (b2);
            \vertex [below=1 of a2] (j1);
            \vertex [sdot] at ($(a2)!1!30:(j1)$) (c1) {};
            \vertex [left=.4cm of c1] (j2);
            \vertex [label=180:$3$] at ($(c1)!1!30:(j2)$) (c0);
            \vertex [right=.4cm of c1] (j3);
            \vertex at ($(c1)!1!30:(j3)$) (c2);
            \diagram*{
                (a0) --[scalar] (a1) --[wavefunc] (a2) --[scalar] (a3),
                (b0) --[scalar] (b1) --[scalar] (b2),
                (c0) --[scalar] (c1) --[scalar] (c2),
                (a1) -- [photon, rmomentum={[compact arrow]$\ \ k_2$}] (b1),
                (a2) -- [photon, rmomentum={[compact arrow]$\ \ k_3$}] (c1),
            };
        \end{feynman}
    \end{tikzpicture}
    =&  {-iy^4} \int \frac{\dd \mu_{1,2,3}}{k_2^2 k_3^2}
    \frac{1}{k_2\ccdot p_1}
    \big( (c_1^{ba} c_2^a c_3^b) (\tilde{c}_1\ccdot \tilde{c}_2) (\tilde{c}_1\ccdot \tilde{c}_3)
    + (c_1\ccdot c_2) (c_1\ccdot c_3) (\tilde{c}_1^{\tilde{b}\tilde{a}} \tilde{c}_2^{\tilde{a}} \tilde{c}_3^{\tilde{b}}) \big)
     \\
    \label{eq:BSNLOoutprop}
    \begin{tikzpicture}[baseline={(current bounding box.center)}]
        \begin{feynman}
            \vertex [label=180:$1$] at (0,0) (a0);
            \vertex [sdot, right=.6cm of a0] (a1) {};
            \vertex [sdot, right=.7cm of a1] (a2) {};
            \vertex [right=.6cm of a2] (a3);
            \vertex [below=1 of a1] (i1);
            \vertex [sdot] at ($(a1)!1!-30:(i1)$) (b1) {};
            \vertex [left=.4cm of b1] (i2);
            \vertex [label=180:$2$] at ($(b1)!1!-30:(i2)$) (b0);
            \vertex [right=.4cm of b1] (i3);
            \vertex at ($(b1)!1!-30:(i3)$) (b2);
            \vertex [below=1 of a2] (j1);
            \vertex [sdot] at ($(a2)!1!30:(j1)$) (c1) {};
            \vertex [left=.4cm of c1] (j2);
            \vertex [label=180:$3$] at ($(c1)!1!30:(j2)$) (c0);
            \vertex [right=.4cm of c1] (j3);
            \vertex at ($(c1)!1!30:(j3)$) (c2);
            \vertex at ($(a2)!0.2!0:(b1)$) (k2) {};
            \vertex at ($(a1)!0.2!(c1)$) (k3) {};
            \diagram*{
                (a0) --[scalar] (a1) --[wavefunc] (a2) --[scalar] (a3),
                (b0) --[scalar] (b1) --[scalar] (b2),
                (c0) --[scalar] (c1) --[scalar] (c2),
                (a1) -- [photon] (c1),
                (k3) -- [opacity=0, rmomentum={[compact arrow]$\ \ k_3$}] (c1),
                (a2) -- [photon] (b1),
                (k2) -- [opacity=0, rmomentum={[compact arrow]$\ \ k_2$}] (b1),
            };
        \end{feynman}
    \end{tikzpicture}
    =& {-iy^4} \int \frac{\dd \mu_{1,2,3}}{k_2^2 k_3^2}
    \frac{-1}{k_2\ccdot p_1}
    \big( (c_1^{ab} c_2^a c_3^b) (\tilde{c}_1\ccdot \tilde{c}_2) (\tilde{c}_1\ccdot \tilde{c}_3)
    + (c_1\ccdot c_2) (c_1\ccdot c_3) (\tilde{c}_1^{\tilde{a}\tilde{b}} \tilde{c}_2^{\tilde{a}} \tilde{c}_3^{\tilde{b}}) \big)
    \\
    \label{eq:BSNLO3phi}
    \begin{tikzpicture}[baseline={(current bounding box.center)}]
        \begin{feynman}
            \vertex [sdot] (g1) {};
            \vertex [below=0.5cm of g1] (g2);
            \vertex [sdot, above=1cm of g1] (a0) {};
            \vertex [sdot, left=0.87cm of g2] (b0) {};
            \vertex [sdot, right=0.87cm of g2] (c0) {};
            \vertex [left=0.6cm of a0, label=180:$1$] (a1);
            \vertex [right=0.6cm of a0] (a2);
            \vertex [above=0.52cm of b0] (i1);
            \vertex [left=0.3cm of i1, label=180:$2$] (b1);
            \vertex [below=0.52cm of b0] (i2);
            \vertex [right=0.3cm of i2] (b2);
            \vertex [below=0.52cm of c0] (i3);
            \vertex [left=0.3cm of i3, label=180:$3$] (c1);
            \vertex [above=0.52cm of c0] (i4);
            \vertex [right=0.3cm of i4] (c2);
            \diagram*{
                (a1) -- [scalar] (a0) -- [scalar] (a2),
                (b1) -- [scalar] (b0) -- [scalar] (b2),
                (c1) -- [scalar] (c0) -- [scalar] (c2),
                (g1) -- [photon, rmomentum={[arrow distance=5pt, label distance=-4pt, arrow shorten=0.2]$k_1$}] (a0),
                (g1) -- [photon, rmomentum={[arrow distance=5pt, label distance=-4pt, arrow shorten=0.2]$k_2$}] (b0),
                (g1) -- [photon, rmomentum={[arrow distance=5pt, label distance=-4pt, arrow shorten=0.2]$k_3$}] (c0),
            };
        \end{feynman}
    \end{tikzpicture}
    =& {-iy^4} \int \frac{\dd \mu_{1,2,3}}{k_1^2 k_2^2 k_3^2}
    \ 2 f^{abc} c_1^a c_2^b c_3^c
    \tilde{f}^{\tilde{a}\tilde{b}\tilde{c}} \tilde{c}_1^{\tilde{a}} \tilde{c}_2^{\tilde{b}} \tilde{c}_3^{\tilde{c}}
\end{align}
\end{widetext}
where for compactness we have defined
\begin{align}
    c^{ab} := \left(\psi^\dagger  T^a T^b \psi \right)\, , \quad
      \tilde {c}^{\tilde a\tilde b} := \left(\tilde\psi^\dagger  \tilde T^{\tilde a} \tilde T^{\tilde b} \tilde\psi \right)\, .
\end{align}
Note that in \eqref{eq:BSNLOinprop} and \eqref{eq:BSNLOoutprop}, the propagator with an arrow denotes either the color or dual color wave function, and we have added up their contributions.
We stress that the factors $c^{ab}$ are absent in the equation of motion, so they will not explicitly appear in the classical solutions \cite{Wong:1970fu}.
In fact, summing up \eqref{eq:BSNLOinprop} and \eqref{eq:BSNLOoutprop} we can remove $c^{ab}$ by
\begin{align}\label{eq:Jacobic}
    c^{ab} - c^{ba} = f^{abc} c^{c}\, ,
\end{align}
and similarly for the dual-color sector.
However, these factors turn out to be critical for the double copy:
because of them we find classical numerators that satisfy color-kinematics duality at this order.
From \eqref{eq:BSNLOzprop} - \eqref{eq:BSNLOoutprop} we can identify 3 (dual-)color factors,
\begin{align}
    \label{eq:eikonalNLOColor}
    C_i^{\mathrm{(123)}} =& \big\{ (c_1\ccdot c_2) (c_1\ccdot c_3) \, ,\,  (c_1^{ab} c_2^a c_3^b) \, ,\,  (c_1^{ba} c_2^a c_3^b)  \big\} \\
    \label{eq:eikonalNLOdColor}
    \tilde{C}_i^{\mathrm{(123)}} =& \big\{ (\tilde{c}_1\ccdot \tilde{c}_2) (\tilde{c}_1\ccdot \tilde{c}_3)\, , \,
    (\tilde{c}_1^{\tilde{a}\tilde{b}} \tilde{c}_2^{\tilde{a}} \tilde{c}_3^{\tilde{b}}) \, , \,
    (\tilde{c}_1^{\tilde{b}\tilde{a}} \tilde{c}_2^{\tilde{a}} \tilde{c}_3^{\tilde{b}})  \big\} \, .
\end{align}
Note that here we only consider diagrams with worldline propagators of particle $1$. There are also contributions involving propagators of $2$ and $3$, which can be gained simply by relabeling $(123)$ in \eqref{eq:BSNLOzprop}-\eqref{eq:BSNLOoutprop} and give us another 6 color factors.
Together with the single (dual)-color factor emerging from \eqref{eq:BSNLO3phi}
\begin{align}
    \label{eq:eikonalNLOColor0}
    C_i^{\mathrm{(0)}} = f^{abc} c_1^a c_2^b c_3^c,
    \qquad
    \tilde{C}_i^{\mathrm{(0)}} = \tilde{f}^{\tilde{a}\tilde{b}\tilde{c}} \tilde{c}_1^{\tilde{a}} \tilde{c}_2^{\tilde{b}} \tilde{c}_3^{\tilde{c}},
\end{align}
we see that the double copy kernel $K_{ij}$ is 10-dimensional.
Fortunately, $K_{ij}$ is block-diagonal. The block that corresponds to the three-dimensional space \eqref{eq:eikonalNLOColor}  is
\begin{align}
    \label{eq:eikonalNLOKernel}
    K_{ij}^{\mathrm{(123)}} =& \frac{1}{k_2^2 k_3^2}  \left(
    \begin{array}{cccc}
        \frac{k_2 \cdot k_3}{(k_2 \cdot p_1)^2}  & \frac{-1}{k_2\cdot p_1} & \frac{1}{k_2\cdot p_1} \\
        \frac{-1}{k_2\cdot p_1} & 0 & 0 \\
        \frac{1}{k_2\cdot p_1} & 0 & 0 \\
    \end{array}
    \right).
\end{align}
and analogously for the color-dual \eqref{eq:eikonalNLOdColor}.
By permutations of $(123)$ we may obtain other blocks.
The last block coupling to the structure constant is extracted from \eqref{eq:BSNLO3phi} and is $1$-dimensional, we have
\begin{gather}
    \label{eq:eikonalNLOKernel0}
    K_{ij}^{\mathrm{(0)}} = \frac{2}{k_1^2 k_2^2 k_3^2}.
\end{gather}

We now proceed to consider the Yang-Mills coupled WQFT (WYM) theory.
The Feynman diagrams are very similar to those of  WBS theory.
With the WYM Feynman rules \eqref{eq:vertex0z1gYM} - \eqref{eq:vertex1psiin1gYM}, we may compute the contributions to the eikonal phase
\begin{widetext}
\begin{align}
    \label{eq:YMNLOzprop}
    &\begin{tikzpicture}[baseline={(current bounding box.center)}]
        \begin{feynman}
            \vertex [label=180:$1$] at (0,0) (a0) ;
            \vertex [sdot, right=.6cm of a0] (a1) {};
            \vertex [sdot, right=.7cm of a1] (a2) {};
            \vertex [right=.6cm of a2] (a3);
            \vertex [below=1 of a1] (i1);
            \vertex [sdot] at ($(a1)!1!-30:(i1)$) (b1) {};
            \vertex [left=.4cm of b1] (i2);
            \vertex [label=180:2] at ($(b1)!1!-30:(i2)$) (b0);
            \vertex [right=.4cm of b1] (i3);
            \vertex at ($(b1)!1!-30:(i3)$) (b2);
            \vertex [below=1 of a2] (j1);
            \vertex [sdot] at ($(a2)!1!30:(j1)$) (c1) {};
            \vertex [left=.4cm of c1] (j2);
            \vertex [label=180:3] at ($(c1)!1!30:(j2)$) (c0);
            \vertex [right=.4cm of c1] (j3);
            \vertex at ($(c1)!1!30:(j3)$) (c2);
            \diagram*{
                (a0) --[scalar] (a1) --[ultra thick] (a2) --[scalar] (a3),
                (b0) --[scalar] (b1) --[scalar] (b2),
                (c0) --[scalar] (c1) --[scalar] (c2),
                (a1) --[gluon, rmomentum={[compact arrow]$\ \ k_2$} ] (b1),
                (a2) --[gluon, rmomentum={[compact arrow]$\ \ k_3$} ] (c1),
            };
        \end{feynman}
    \end{tikzpicture}
    = {-ig^4} \int \frac{\dd \mu_{1,2,3}}{k_2^2 k_3^2}
     (c_1\ccdot c_2) (c_1\ccdot c_3)
     \left( \frac{k_2\ccdot k_3}{(k_2\ccdot p_1)^2}  n_0
     + \frac{ 1 }{k_2 \ccdot p_1} n_1
    \right) \\
    \label{eq:YMNLOinprop}
    &\begin{tikzpicture}[baseline={(current bounding box.center)}]
        \begin{feynman}
            \vertex [label=180:$1$] at (0,0) (a0) ;
            \vertex [sdot, right=.6cm of a0] (a1) {};
            \vertex [sdot, right=.7cm of a1] (a2) {};
            \vertex [right=.6cm of a2] (a3);
            \vertex [below=1 of a1] (i1);
            \vertex [sdot] at ($(a1)!1!-30:(i1)$) (b1) {};
            \vertex [left=.4cm of b1] (i2);
            \vertex [label=180:2] at ($(b1)!1!-30:(i2)$) (b0);
            \vertex [right=.4cm of b1] (i3);
            \vertex at ($(b1)!1!-30:(i3)$) (b2);
            \vertex [below=1 of a2] (j1);
            \vertex [sdot] at ($(a2)!1!30:(j1)$) (c1) {};
            \vertex [left=.4cm of c1] (j2);
            \vertex [label=180:3] at ($(c1)!1!30:(j2)$) (c0);
            \vertex [right=.4cm of c1] (j3);
            \vertex at ($(c1)!1!30:(j3)$) (c2);
            \diagram*{
                (a0) --[scalar] (a1) --[wavefunc] (a2) --[scalar] (a3),
                (b0) --[scalar] (b1) --[scalar] (b2),
                (c0) --[scalar] (c1) --[scalar] (c2),
                (a1) --[gluon, rmomentum={[compact arrow]$\ \ k_2$} ] (b1),
                (a2) --[gluon, rmomentum={[compact arrow]$\ \ k_3$} ] (c1),
            };
        \end{feynman}
    \end{tikzpicture}
    = {-ig^4} \int \frac{\dd \mu_{1,2,3}}{k_2^2 k_3^2}
    \frac{(c_1^{ba} c_2^a c_3^b) }{k_2\ccdot p_1}  n_0 \\
    \label{eq:YMNLOoutprop}
    &\begin{tikzpicture}[baseline={(current bounding box.center)}]
        \begin{feynman}
            \vertex [label=180:$1$] at (0,0) (a0);
            \vertex [sdot, right=.6cm of a0] (a1) {};
            \vertex [sdot, right=.7cm of a1] (a2) {};
            \vertex [right=.6cm of a2] (a3);
            \vertex [below=1 of a1] (i1);
            \vertex [sdot] at ($(a1)!1!-30:(i1)$) (b1) {};
            \vertex [left=.4cm of b1] (i2);
            \vertex [label=180:$2$] at ($(b1)!1!-30:(i2)$) (b0);
            \vertex [right=.4cm of b1] (i3);
            \vertex at ($(b1)!1!-30:(i3)$) (b2);
            \vertex [below=1 of a2] (j1);
            \vertex [sdot] at ($(a2)!1!30:(j1)$) (c1) {};
            \vertex [left=.4cm of c1] (j2);
            \vertex [label=180:$3$] at ($(c1)!1!30:(j2)$) (c0);
            \vertex [right=.4cm of c1] (j3);
            \vertex at ($(c1)!1!30:(j3)$) (c2);
            \vertex at ($(a2)!0.2!0:(b1)$) (k2) {};
            \vertex at ($(a1)!0.2!(c1)$) (k3) {};
            \diagram*{
                (a0) --[scalar] (a1) --[wavefunc] (a2) --[scalar] (a3),
                (b0) --[scalar] (b1) --[scalar] (b2),
                (c0) --[scalar] (c1) --[scalar] (c2),
                (a1) -- [gluon] (c1),
                (k3) -- [opacity=0, rmomentum={[compact arrow]$\ \ k_3$}] (c1),
                (a2) -- [gluon] (b1),
                (k2) -- [opacity=0, rmomentum={[compact arrow]$\ \ k_2$}] (b1),
            };
        \end{feynman}
    \end{tikzpicture}
    = {-ig^4} \int \frac{\dd \mu_{1,2,3}}{k_2^2 k_3^2}
    \frac{-(c_1^{ab} c_2^a c_3^b) }{k_2\ccdot p_1} n_0 \\
    \label{eq:YMNLO3g}
    &    \begin{tikzpicture}[baseline={(current bounding box.center)}]
        \begin{feynman}
            \vertex [sdot] (g1) {};
            \vertex [below=0.5cm of g1] (g2);
            \vertex [sdot, above=1cm of g1] (a0) {};
            \vertex [sdot, left=0.87cm of g2] (b0) {};
            \vertex [sdot, right=0.87cm of g2] (c0) {};
            \vertex [left=0.6cm of a0, label=180:$1$] (a1);
            \vertex [right=0.6cm of a0] (a2);
            \vertex [above=0.52cm of b0] (i1);
            \vertex [left=0.3cm of i1, label=180:$2$] (b1);
            \vertex [below=0.52cm of b0] (i2);
            \vertex [right=0.3cm of i2] (b2);
            \vertex [below=0.52cm of c0] (i3);
            \vertex [left=0.3cm of i3, label=180:$3$] (c1);
            \vertex [above=0.52cm of c0] (i4);
            \vertex [right=0.3cm of i4] (c2);
            \diagram*{
                (a1) -- [scalar] (a0) -- [scalar] (a2),
                (b1) -- [scalar] (b0) -- [scalar] (b2),
                (c1) -- [scalar] (c0) -- [scalar] (c2),
                (g1) -- [gluon, rmomentum={[arrow distance=5pt, label distance=-4pt, arrow shorten=0.2]$k_1$}] (a0),
                (g1) -- [gluon, rmomentum={[arrow distance=5pt, label distance=-4pt, arrow shorten=0.2]$k_2$}] (b0),
                (g1) -- [gluon, rmomentum={[arrow distance=5pt, label distance=-4pt, arrow shorten=0.2]$k_3$}] (c0),
            };
        \end{feynman}
    \end{tikzpicture}
    = {-ig^4} \int \dd \mu_{1,2,3}
    \frac{2 f^{abc} c_1^a c_2^b c_3^c} {k_1^2 k_2^2 k_3^2} \left( -n_1 \right)
\end{align}
\end{widetext}
where we have defined
\begin{align}
    &n_0 = p_1\ccdot p_2\, p_1\ccdot p_3 \\
    &n_1 = k_2 \ccdot p_3\, p_1 \ccdot p_2 - k_3 \ccdot p_2\, p_1 \ccdot p_3 - k_2 \ccdot p_1\, p_2 \ccdot p_3.
\end{align}
Based on the color factors identified in \eqref{eq:eikonalNLOColor}, \eqref{eq:eikonalNLOColor0} and the double copy kernel \eqref{eq:eikonalNLOKernel}, \eqref{eq:eikonalNLOKernel0}, we are led to organize the numerators as
\begin{align}
    \label{eq:N123atNLO}
    N_j^{\mathrm{(123)}} =& \left\{ n_0 \, ,\,  \frac{-n_1}{2} \,,  \, \frac{n_1}{2} \right\} \\
    \label{eq:N0atNLO}
    N_j^{\mathrm{(0)}} =& \, -n_1 \, ,
\end{align}
so that the WYM eikonal may be decomposed in the form of \eqref{eq:dcEikonalYM},
\begin{align}
    \chi_2 =
     -g^4 &\int \dd \mu_{1,2,3}
    \sum_{i,j} \Big(
        C_i^{\mathrm{(0)}} K_{ij}^{\mathrm{(0)}} N_j^{\mathrm{(0)}} \nonumber\\
        &+ \big(C_i^{\mathrm{(123)}} K_{ij}^{\mathrm{(123)}} N_j^{\mathrm{(123)}} + \text{cyclic} \big)
    \Big).
\end{align}
Fortunately, this decomposition automatically satisfies the color-kinematics duality
\begin{align}
    c_1^{ab} c_2^a c_3^b - c_1^{ba} c_2^a c_3^b =& f^{abc} c_1^a c_2^b c_3^c \\
     \frac{-n_1}{2} - \frac{n_1}{2} =& -n_1.
\end{align}
Note that the decomposition of $N_j^{(123)}$ is not unique due to the Jacobi relation \eqn{eq:Jacobic}\footnote{For example, we could also have $N_j^{\mathrm{(123)}} = \left( n_0 \quad {0} \quad {n_1} \right)$. Color-kinematic duality still holds and the double copy gives the correct gravitational result. We have chosen to write $N_j^{(123)}$ in a symmetric form.}.

In principal, we are now prepared to execute the double copy as proposed in \eqref{eq:dcEikonalDG} to get the eikonal phase in the worldline coupled dilaton gravity theory (WDG).
In order to check the validity of our double copy prescription,
we directly compute the eikonal in WDG theory with \eqref{eq:vertex0z1hDG} - \eqref{eq:vertex2z1hDG} we find for the graphs
not involving bulk graviton interactions
\begin{widetext}
\begin{align}
    \label{eq:DGNLO1z}
    \begin{tikzpicture}[baseline={(0,-0.7)}]
        \begin{feynman}
            \vertex [label=180:$1$] at (0,0) (a0) ;
            \vertex [sdot, right=.6cm of a0] (a1) {};
            \vertex [sdot, right=.7cm of a1] (a2) {};
            \vertex [right=.6cm of a2] (a3);
            \vertex [below=1 of a1] (i1);
            \vertex [sdot] at ($(a1)!1!-30:(i1)$) (b1) {};
            \vertex [left=.4cm of b1] (i2);
            \vertex [label=180:2] at ($(b1)!1!-30:(i2)$) (b0);
            \vertex [right=.4cm of b1] (i3);
            \vertex at ($(b1)!1!-30:(i3)$) (b2);
            \vertex [below=1 of a2] (j1);
            \vertex [sdot] at ($(a2)!1!30:(j1)$) (c1) {};
            \vertex [left=.4cm of c1] (j2);
            \vertex [label=180:3] at ($(c1)!1!30:(j2)$) (c0);
            \vertex [right=.4cm of c1] (j3);
            \vertex at ($(c1)!1!30:(j3)$) (c2);
            \diagram*{
                (a0) --[scalar] (a1) --[ultra thick] (a2) --[scalar] (a3),
                (b0) --[scalar] (b1) --[scalar] (b2),
                (c0) --[scalar] (c1) --[scalar] (c2),
                (a1) --[graviton] (b1),
                (a1) --[opacity=0, rmomentum={[compact arrow]$\ \ k_2$} ] (b1),
                (a2) --[graviton] (c1),
                (a2) --[opacity=0, rmomentum={[compact arrow]$\ \ k_3$} ] (c1),
            };
        \end{feynman}
    \end{tikzpicture}
    &
    \begin{aligned}
    = \frac{-i\kappa^4}{16} \int& \frac{\dd \mu_{1,2,3}} {k_2^2 k_3^2 (k_2\ccdot p_1)^2}
    \big( (k_2\ccdot k_3)  (p_1\ccdot p_2)^2 (p_1\ccdot p_3)^2
    -4 (k_2\ccdot p_1)^2 (p_1\ccdot p_2) (p_1\ccdot p_3) (p_2\ccdot p_3) \\
    & \quad
    -2 (k_3\ccdot p_2) (k_2 \ccdot p_1) (p_1\ccdot p_2) (p_1\ccdot p_3)^2
    +2 (k_2\ccdot p_3) (k_2 \ccdot p_1) (p_1\ccdot p_3) (p_1\ccdot p_2)^2
    \big)
    \end{aligned}\\
    \label{eq:DGNLO0z}
    \begin{tikzpicture}[baseline={(0,-0.7)}]
        \begin{feynman}
            \vertex [label=180:$1$] at (0,0) (a0) ;
            \vertex [sdot, right=.8cm of a0] (a1) {};
            \vertex [right=.8cm of a1] (a3);
            \vertex [below=1 of a1] (i1);
            \vertex [sdot] at ($(a1)!1!-40:(i1)$) (b1) {};
            \vertex [left=.4cm of b1] (i2);
            \vertex [label=180:2] at ($(b1)!1!-40:(i2)$) (b0);
            \vertex [right=.4cm of b1] (i3);
            \vertex at ($(b1)!1!-40:(i3)$) (b2);
            \vertex [sdot] at ($(a1)!1!40:(i1)$) (c1) {};
            \vertex [left=.4cm of c1] (j2);
            \vertex at ($(c1)!1!40:(j2)$) (c0);
            \vertex [right=.4cm of c1] (j3);
            \vertex [label=0:3] at ($(c1)!1!40:(j3)$) (c2);
            \diagram*{
                (a0) --[scalar] (a1) --[scalar] (a3),
                (b0) --[scalar] (b1) --[scalar] (b2),
                (c0) --[scalar] (c1) --[scalar] (c2),
                (a1) --[graviton] (b1),
                (a1) --[opacity=0, rmomentum'={[compact arrow]$k_2\ \ $} ] (b1),
                (a1) --[graviton] (c1),
                (a1) --[opacity=0, rmomentum={[compact arrow]$\ \ k_3$} ] (c1),
            };
        \end{feynman}
    \end{tikzpicture}
    &= \frac{-i\kappa^4}{16} \int \frac{\dd \mu_{1,2,3}}{k_2^2 k_3^2}
    \big(2 (p_1\ccdot p_2) (p_1\ccdot p_3) (p_2\ccdot p_3) \big).
\end{align}
\end{widetext}
Summing up the two diagrams \eqref{eq:DGNLO1z} and \eqref{eq:DGNLO0z}, we can check that the result can be written as
\begin{align}
    \label{eq:dcNLO1}
     & \frac{-i\kappa^4}{16} \int \frac{\dd \mu_{1,2,3}}{k_2^2 k_3^2}
    \left(  \frac{k_2\ccdot k_3 n_0^2}{(k_2\ccdot p_1)^2}
    + \frac{2 n_0 n_1}{k_2\ccdot p_1} \right)  \\
    =& \frac{-i\kappa^4}{16} \int {\dd \mu_{1,2,3}}
     \sum_{i,j} N_i^{\mathrm{(123)}} K_{ij}^{\mathrm{(123)}} N_j^{\mathrm{(123)}}. \nonumber
\end{align}
In the last line we have arranged the result to the form of \eqref{eq:dcEikonalDG} with the double copy kernel $K_{ij}^{\mathrm{(123)}}$ and the numerator $N_i^{\mathrm{(123)}}$ defined in \eqref{eq:eikonalNLOKernel} and \eqref{eq:N123atNLO} respectively.
Turning to the bulk graviton interaction graphs
thanks to the field redefinition of $\{\varphi, h_{\mu\nu}\}$, the three-graviton vertex \eqref{eq:vertex3h} is directly proportional to the square of three-gluon vertex, so we can easily compute the last diagram which is manifestly a double-copy of the WYM one
\begin{align}
    \label{eq:dcNLO2}
    &\begin{tikzpicture}[baseline={(current bounding box.center)}]
        \begin{feynman}
            \vertex [sdot] (g1) {};
            \vertex [below=0.5cm of g1] (g2);
            \vertex [sdot, above=1cm of g1] (a0) {};
            \vertex [sdot, left=0.87cm of g2] (b0) {};
            \vertex [sdot, right=0.87cm of g2] (c0) {};
            \vertex [left=0.6cm of a0, label=180:$1$] (a1);
            \vertex [right=0.6cm of a0] (a2);
            \vertex [above=0.52cm of b0] (i1);
            \vertex [left=0.3cm of i1, label=180:$2$] (b1);
            \vertex [below=0.52cm of b0] (i2);
            \vertex [right=0.3cm of i2] (b2);
            \vertex [below=0.52cm of c0] (i3);
            \vertex [left=0.3cm of i3, label=180:$3$] (c1);
            \vertex [above=0.52cm of c0] (i4);
            \vertex [right=0.3cm of i4] (c2);
            \diagram*{
                (a1) -- [scalar] (a0) -- [scalar] (a2),
                (b1) -- [scalar] (b0) -- [scalar] (b2),
                (c1) -- [scalar] (c0) -- [scalar] (c2),
                (g1) -- [graviton] (a0),
                (g1) -- [graviton] (b0),
                (g1) -- [graviton] (c0),
                (g1) -- [opacity=0, rmomentum={[arrow distance=5pt, label distance=-4pt, arrow shorten=0.2]$k_1$}] (a0),
                (g1) -- [opacity=0, rmomentum={[arrow distance=5pt, label distance=-4pt, arrow shorten=0.2]$k_2$}] (b0),
                (g1) -- [opacity=0, rmomentum={[arrow distance=5pt, label distance=-4pt, arrow shorten=0.2]$k_3$}] (c0),
            };
        \end{feynman}
    \end{tikzpicture}
    = \frac{-i\kappa^4}{16} \int \dd \mu_{1,2,3} \frac{2 n_1^2}{k_1^2 k_2^2 k_3^2}.
\end{align}
From \eqref{eq:dcNLO1} and \eqref{eq:dcNLO2} we therefore conclude that the double copy of the WYM eikonal coincides with the one of WDG  also at the next-to-leading order ($\mathcal{O}(\kappa^4)$).

\section{Radiative double copy}
\label{sec3}
In this letter we are mainly considering the conservative sector of the WQFT, however, with a slight modification we can generalize the eikonal double copy \eqref{eq:dcEikonal} to classical radiation.
In WQFT, the $\Phi$ field radiation is computed as \cite{Jakobsen:2021smu,Jakobsen:2021lvp}
\begin{align}
    -i k^2 \left. \langle \Phi(k) \rangle \right|_{k^2=0}
\end{align}
For $\Phi \in \{A_\mu^a, h_{\mu\nu} \}$, we also need to contract it with the polarizations $\{\epsilon^\mu, \epsilon^{\mu\nu}\}$ respectively.
We take the gluon radiation as an example.
Loosely speaking, the radiation at order $\mathcal{O}(g^{2n-1})$ can be obtained from the eikonal phase at $\mathcal{O}(g^{2n})$ by cutting off one worldline.
Diagrammatically, the gluon radiation of a binary source at leading order can be gained from \eqref{eq:YMNLOzprop}-\eqref{eq:YMNLO3g} by cutting the propagator $k_3$ and identifying $k_3$ with the momentum of the radiated gluon.
The on-shell condition $k_3 \cdot \epsilon = 0$ plays the same role as the $\ddelta(k_3 \cdot p_3)$ in the measure of the eikonal phase.
This ensures that the gluon radiation can be decomposed into $C_i K_{ij} N_j$, with $C_i$ attained from \eqref{eq:eikonalNLOColor} and \eqref{eq:BSNLO3phi} by striping off $c_3$, $N_i$ from \eqref{eq:N123atNLO} and \eqref{eq:N0atNLO} by replacing $p_3^\mu$ by $\epsilon^\mu$ and $K_{ij}$ being identical to \eqref{eq:eikonalNLOKernel}.
From the same approach we can also get the gravitational radiation and decompose it as $N_i K_{ij} N_j$.
Therefore we conclude that the double copy construction works for radiation, too.
We note that this is equivalent to the approach considered by Shen \cite{Shen:2018ebu} and Goldberger and Ridgway \cite{Goldberger:2016iau} where the radiation is calculated by solving the equations of motion.

\section{From amplitude to eikonal}
\label{sec4}
The expectation values in WQFT are directly linked to the classical limit of S-matrix element.
Consequently, we can expect that the classical double copy of WQFT discussed is also closely related to the double copy at the level of the scattering amplitude.
In this section, we will consider scalar QCD, i.e.~massive scalar fields coupled to Yang-Mills whose double copy has been studied in \cite{Plefka:2019wyg}.
We claim that the classical limit of the scattering amplitude of $n$ distinct scalar pairs corresponds to the WYM eikonal phase at $\mathcal{O} \left(g^{2(n-1)} \right)$ and show the connection explicitly at  $\mathcal{O} \left(g^{4} \right)$.
Moreover, we will demonstrate that the double copy of the eikonal phase is the classical limit of the BCJ double copy of the scattering amplitude.

The exponentiated eikonal phase is directly related to the classical limit of scattering amplitude\cite{Amati:1987wq, Amati:1990xe},
\begin{align}
(1+\Delta_{q})    e^{i \chi} -1 = \sum_{n=2} \frac{1}{2^n} \int {\dd \mu_{1,2,...,n}} \lim_{\hbar \to 0}\mathcal{A} (n \to n)
\end{align}
where $\mathcal{A} (n \to n)$ denotes an amplitude of $n$ pairs of distinct massive scalars,
and $\chi$ is the total eikonal phase, which scales as $\hbar^{-1}$ and receives contributions from all higher-loop amplitudes.
The introduction of the  ``quantum remainder'' $\Delta_{q}$ (scaling as $\hbar^{n\geq 0}$) is needed for consistency \cite{Heissenberg:2021tzo}.
Here, we only care about tree diagrams, therefore we have
\begin{align}
    \label{eq:AtoEikonal}
    \chi_{n-1} = \frac{-i}{2^n} \int {\dd \mu_{1,2,...,n}} \lim_{\hbar \to 0}\mathcal{A}^{\mathrm{tree}} (n \to n).
\end{align}
The correspondence at the $2\to 2$ level is rather trivial, so we will focus on the $3\to 3$ case.
The leading order 6-scalar amplitude in SQCD is  \cite{Plefka:2019wyg}\footnote{We have converted the result of \cite{Plefka:2019wyg} to follow our conventions.}
\begin{align} \label{eq:6scalars}
        \mathcal{A}^{\mathrm{tree}} (3& \to 3) =
        \begin{tikzpicture}[baseline={(current bounding box.center)}]
        \begin{feynman}
            \vertex [small blob] (c) {};
            \vertex [left=0.9cm of c, label=180:\(\hat{p}_2\!+\!\frac{k_2}{2}\text{, } l\)] (l2);
            \vertex [label=180:$\hat{p}_1\!+\!\frac{k_1}{2}\text{, } j$] at ($(c)!1!-60:(l2)$) (l1);
            \vertex [label=180:$\hat{p}_3\!+\!\frac{k_3}{2}\text{, } n$] at ($(c)!1!60:(l2)$) (l3);
            \vertex [right=0.9cm of c, label=0:$k\text{, }\hat{p}_2\!-\!\frac{k_2}{2}$] (r2);
            \vertex [label=0:$i\text{, }\hat{p}_1\!-\!\frac{k_1}{2}$] at ($(c)!1!60:(r2)$) (r1);
            \vertex [label=0:$m\text{, }\hat{p}_3\!-\!\frac{k_3}{2}$] at ($(c)!1!-60:(r2)$) (r3);
            \diagram*{
                (l1) -- [fermion, arrow size=1.1pt, red] (c) -- [fermion, arrow size=1.1pt, red] (r1),
                (l2) -- [fermion, arrow size=1.1pt, blue] (c) -- [fermion, arrow size=1.1pt, blue] (r2),
                (l3) -- [fermion, arrow size=1.1pt, green] (c) -- [fermion, arrow size=1.1pt, green] (r3),
            };
        \end{feynman}
    \end{tikzpicture}
    \nonumber \\
    =& {8\hat{c}^{(0)} \hat{n}^{(0)} \over k_1^2 k_2^2 k_3^2}
    + \bigg[ \frac{8}{k_2^2 k_3^2} \bigg(
    \frac{\hat{c}^{(123)} \hat{n}^{(123)}}{2 \hat{p}_1\ccdot k_2 - k_2 \ccdot k_3} \nonumber\\
    & + \frac{\hat{c}^{(132)} \hat{n}^{(132)}}{2 \hat{p}_1\ccdot k_3 - k_3 \ccdot k_2}
    \bigg) + \text{cyclic} \bigg],
\end{align}
where we have introduced $\hat{p}_{i}$ as the average of the in- and outgoing momentum of particle $i$ which is orthogonal
to its momentum transfer $\hat{p}_i \cdot k_i =0$.
The color factors are
\begin{align}
    \hat{c}^{(0)} &= f^{abc} T^a_{ij} T^b_{kl} T^c_{mn} \nonumber \\
    \hat{c}^{(123)} &= (T^b T^a)_{ij} T^a_{kl} T^b_{mn} \\
    \hat{c}^{(132)} &= (T^a T^b)_{ij} T^a_{kl} T^b_{mn}.\nonumber
\end{align}
where $i,j,...,l$ denote the color indices of the massive scalars.
The corresponding numerators are
\begin{align}
    \hat{n}^{(0)} =& {-ig^4} \hat{p}_{1,\mu} \hat{p}_{2,\nu} \hat{p}_{3,\rho}  V^{\mu\nu\rho}_{123} \\
    \label{eq:numerator123}
    \hat{n}^{(123)}\! =& \frac{-ig^4}{2} \Big( 4\hat{p}_1\ccdot \hat{p}_2\, \hat{p}_1\ccdot \hat{p}_3
     + 2\hat{p}_1\ccdot \hat{p}_3 \, k_1\ccdot \hat{p}_2
     - 2 \hat{p}_1\ccdot \hat{p}_2 \, k_1\ccdot \hat{p}_3
     \nonumber \\ &
     - 2 \hat{p}_1 \ccdot k_2 \, \hat{p}_2 \ccdot \hat{p}_3
     - k_1 \ccdot \hat{p}_2 \, k_1\ccdot \hat{p}_3
     + k_2 \ccdot k_3 \, \hat{p}_2\ccdot \hat{p}_3 \Big) \\
     \label{eq:numerator132}
    \hat{n}^{(132)}\! =& \frac{-ig^4}{2} \Big( 4\hat{p}_1\ccdot \hat{p}_2\, \hat{p}_1\ccdot \hat{p}_3
     + 2\hat{p}_1\ccdot \hat{p}_2 \, k_1\ccdot \hat{p}_3
     - 2 \hat{p}_1\ccdot \hat{p}_3 \, k_1\ccdot \hat{p}_2
     \nonumber \\ &
     - 2 \hat{p}_1 \ccdot k_3 \, \hat{p}_2 \ccdot \hat{p}_3
     - k_1 \ccdot \hat{p}_2 \, k_1\ccdot \hat{p}_3
     + k_2 \ccdot k_3 \, \hat{p}_2\ccdot \hat{p}_3 \Big),
\end{align}
which has been brought into a form to satisfy color-kinematic duality $\hat{n}^{(132)} - \hat{n}^{(123)} = \hat{n}^{(0)}$.
In the classical limit we take small momentum transfers $k_i \to \hbar k_i$, and consider the expansion in small $\hbar$ following \cite{Kosower:2018adc}.
In \eqref{eq:numerator123} and \eqref{eq:numerator132}, we have already sorted the terms in  powers of $k_i$.
We identify the momentum as $\hat{p}_i = p_i$, although since $\hat{p}_i ^2 \neq m_i^2$, we need to change the definition of $p_i$ to $p_i=\hat{m}_i v_i$ with $\hat{m}_i^2 = \hat{p}_i ^2$.\footnote{This is related to the fact that we use Feynman propagators in the amplitudes, for the details, see \cite{Mogull:2020sak}.}
At this order, the redefinition will not change the WQFT result.
The massive propagators will become
\begin{align}
    \frac{1}{2 \hat{p}_1\ccdot k_2 - k_2 \ccdot k_3} \to
    \frac{1}{\hbar} \frac{1}{2 p_1\ccdot k_2} + \frac{k_2 \ccdot k_3}{4 (p_1\ccdot k_2)^2}
    + \mathcal{O}(\hbar).
\end{align}
Performing the classical limit of the Yang-Mills amplitude, we also need to consider the classical limit of the color factors, which was recently investigated by de la Cruz et al. \cite{delaCruz:2020bbn}.
Built on their insight, we propose the classical limit of the color factors to be
\begin{align}
    T^a_{ij} \rightarrow&\  c^a \\
    \label{eq:classicalTT}
    (T^aT^b)_{ij} \rightarrow&\  c^a c^b + \hbar\, c^{ab}\\
    f^{abc} \rightarrow&\  \hbar f^{abc}.
\end{align}
Note that the sub-leading term in \eqref{eq:classicalTT} guarantees that the Jacobi identity holds in the classical limit.

It is now straightforward to compute the the classical limit of the amplitude \eqref{eq:6scalars} and extract the eikonal using \eqref{eq:AtoEikonal}. Keeping only the leading order terms in the classical $\hbar \to 0$ limit, we have
\begin{align}
    {\hat{c}^{(0)} \hat{n}^{(0)} \over k_1^2 k_2^2 k_3^2}
    \to&\  C_i^{(0)} K_{ij}^{(0)} N_j^{(0)} \\
    \frac{1}{ k_2^2 k_3^2 } \bigg(
    \frac{\hat{c}^{(123)} \hat{n}^{(123)}}{ 2 \hat{p}_1\ccdot k_2 - k_2 \ccdot k_3 } +& \frac{\hat{c}^{(132)} \hat{n}^{(132)}}{ 2 \hat{p}_1\ccdot k_3 - k_2 \ccdot k_3 } \bigg)
    \nonumber\\
    \to& \  C_i^{(123)} K_{ij}^{(123)} N_j^{(123)}.
\end{align}
We therefore recover the eikonal phase of SQCD from the WQFT, which directly operates at the classical level.

We can double copy the SQCD amplitude \eqref{eq:6scalars} by replacing the color factors by the numerators.
This results in an amplitude of massive scalars coupled to gravity and the dilaton.
We can then likewise consider the classical limit of this gravitational amplitude,
\begin{align}
    {\hat{n}^{(0)} \hat{n}^{(0)} \over k_1^2 k_2^2 k_3^2}
    \to&\  N_i^{(0)} K_{ij}^{(0)} N_j^{(0)} \\
    \frac{1}{ k_2^2 k_3^2 } \bigg(
    \frac{\hat{n}^{(123)} \hat{n}^{(123)}}{ 2 \hat{p}_1\ccdot k_2 - k_2 \ccdot k_3 } +& \frac{\hat{n}^{(132)} \hat{n}^{(132)}}{ 2 \hat{p}_1\ccdot k_3 - k_2 \ccdot k_3 } \bigg)
    \nonumber\\
    \to& \  N_i^{(123)} K_{ij}^{(123)} N_j^{(123)},
\end{align}
which coincides with our calculation in WDG.
We have therefore verified that the classical double copy of the world line quantum field theory is in full agreement with the quantum double copy of amplitudes at LO and NLO.
Note that the double copy of SQCD contains self-interactions of massive scalars \cite{Plefka:2019wyg}, however, these are short-range interactions and do not contribute to the classical theory. So we don't need to introduce additional terms in WDG and the double copy automatically works out.

\section{Conclusions}

In this work, we have extended the WQFT formalism to massive, charged point particles coupled to the bi-adjoint scalar field, Yang-Mills and dilaton-gravity theories.
We proposed a classical double copy prescription for the eikonal phases, alias free energies, in these theories and explicitly
verified the validity of the double copy up to quartic order (NLO) in the coupling constants. This entails the double copy relation
for the particle's deflection (or scattering angle) as the derivative of the eikonal with respect to the impact parameter.
With minor modifications our double copy prescription also applies to classical radiation emitted in the scattering process.
As a technical tool it was necessary to increase the number of worldlines of scattered particles with the order of
perturbation theory, i.e.~an $(n+2)$-body system for the N${}^{n}$LO order as well as to consider the eikonal of WBS in parallel
to establish the double copy kernels.
In fact, we expect all expectation values in WYM and WDG to feature the double copy relation as they are directly related to the quantum scattering amplitudes.
To illustrate the connection, we compared our eikonal phase to the classical limit of the corresponding eikonal emerging from the massive scalar six-point amplitude finding agreement.

These insights give us the expectation that the classical double copy for the  WQFT will prevail to NNLO and higher. This
would cure the breakdown observed in \cite{Plefka:2019hmz} of a double copy prescription for the off-shell effective action of the
particle's worldline coordinates $x^{\mu}(\tau)$. The essential difference of our approach to the  off-shell effective action of \cite{Plefka:2018dpa,Plefka:2019hmz} is that in WQFT \emph{both} the force
mediating fields and the fluctuations on the worldline are integrated out in the path integral.

The most important application of WQFT will be to Einstein gravity. This, however, continues to be a challenge for the
double copy prescription as it suffers from pollution of the dilaton and in principle even the Kalb-Ramond two-form.
Many approaches have been explored to remove the dilaton from the double copy construction
\cite{Luna:2017dtq,Johansson:2014zca,Bern:2019crd,Carrasco:2021bmu}.
Since the WQFT provides a simple way to extract classical quantities, it might be easier to project out dilaton
at the classical level. It will be interesting to explore this in future work.

An obvious extension of the classical double copy is to incorporate spin, some attempts in this direction are \cite{Goldberger:2017ogt,Li:2018qap, Goldberger:2019xef}.
The recently discovered hidden supersymmetry in the worldline description of spinning compact bodies \cite{Jakobsen:2021zvh}
should be applicable to the Yang-Mills case as well. In particular the limitation of coupling higher-spin worldline theories
to gravity might be overcome upon using the double copy. Concretely, the construction of the $\sqrt{\text{Kerr}}$ solution
\cite{Arkani-Hamed:2019ymq} in the language of the WQFT would be an interesting starting point.

\acknowledgments
This project has received funding from the European Union's Horizon 2020 research
and innovation program under the Marie Sklodowska-Curie grant agreement No. 764850 ``SAGEX''.
JP thanks the Max-Planck-Institut f\"ur Physik (Werner-Heisenberg Institut) for hospitality.
Some of our figures were produced with the help of TikZ-Feynman \cite{Ellis:2016jkw}.

\appendix
\section{Convention} \label{Ap:convention}
We use the mostly minus signature for the spacetime metric.
The generators and structure constants of the color gauge group are normalized such that
\begin{gather}
    \left[T^{a}, T^{b}\right] = f^{a b c} T^{c} \\
    \operatorname{Tr} \left(T^{a} T^{b}\right) = \frac{\delta^{a b}}{2}.
\end{gather}
Note that our $f^{abc}$ is different from a more usual convention by a factor of $i$.
For the YM theory, we use the standard Lagrangian $S^{\mathrm{YM}} = - \int \dd^4x (F^a_{\mu\nu})^2/4$, where
\begin{align}
    F^a_{\mu\nu} = \partial_\mu A_\nu^a - \partial_\nu A_\mu^a - i g f^{abc} A_\mu^b A_\nu^c.
\end{align}
In the Feynman gauge, the gluon propagator and the three-gluon vertex are
\begin{align}
    \label{eq:propA}
    \begin{tikzpicture}[baseline={(0,0)}]
    \begin{feynman}
        \vertex at (-1.85,0) (v1) {$A^a_\mu$};
        \vertex at (0,0) (v2) {$A^b_\nu$};
        \diagram*{
            (v1) -- [gluon, momentum={$k$}] (v2)
        };
    \end{feynman}
    \end{tikzpicture}
    =& \frac{-i}{k^2} \eta_{\mu\nu} \delta^{ab} \\
    \label{eq:vertex3h}
    \begin{tikzpicture}[baseline={(current bounding box.center)}]
    \begin{feynman}
        \vertex [sdot] (g1) {};
        \vertex [above=1.3cm of g1] (a0) {$A^a_\mu$};
        \vertex at ($(g1)!1!120:(a0)$) (b0) {$A^b_\nu$};
        \vertex at ($(g1)!1!120:(b0)$) (c0) {$A^c_\rho$};
        \diagram*{
            (g1) -- [gluon, momentum={[arrow distance=5pt, label distance=-4pt, arrow shorten=0.2]$k_1$}] (a0),
            (g1) -- [gluon, momentum={[arrow distance=5pt, label distance=-4pt, arrow shorten=0.2]$k_2$}] (b0),
            (g1) -- [gluon, momentum={[arrow distance=5pt, label distance=-4pt, arrow shorten=0.2]$k_3$}] (c0),
        };
    \end{feynman}
    \end{tikzpicture}
    =& i g f^{abc} V_{123}^{\mu\nu\rho}
\end{align}
where
\begin{align}
    V_{123}^{\mu\nu\rho} = \big[\eta^{\mu\nu}(k_1 \!-\! k_2)^\rho
    +\eta^{\nu\rho}(k_2 \!-\! k_3)^\mu +\eta^{\rho\mu}(k_3 \!-\! k_1)^\nu \big].
\end{align}

The action of bi-adjoint scalar theory is
\begin{align}
    S^{\mathrm{BS}} \!=\! \int \dd^{4} x\left(\frac{1}{2}\left(\partial_{\mu} \phi^{a \tilde{a}}\right)^{2} \!-\! \frac{y}{3} f^{a b c} \tilde{f}^{\tilde{a} \tilde{b} \tilde{c}} \phi_{a \tilde{a}} \phi_{b \tilde{b}} \phi_{c \tilde{c}}\right)
\end{align}
The Feynman rules are
\begin{align}
    \begin{tikzpicture}[baseline={(0,0)}]
    \begin{feynman}
        \vertex at (-1.85,0) (v1) {$\phi^{a\tilde{a}}$};
        \vertex at (0,0) (v2) {$\phi^{b\tilde{b}}$};
        \diagram*{
            (v1) -- [photon, momentum={$k$}] (v2)
        };
    \end{feynman}
    \end{tikzpicture}
    =& \frac{i}{k^2} \\
    \label{eq:vertex3phi}
    \begin{tikzpicture}[baseline={(current bounding box.center)}]
    \begin{feynman}
        \vertex [sdot] (g1) {};
        \vertex [above=1.0cm of g1] (a0) {$\phi^{a\tilde{a}}$};
        \vertex at ($(g1)!1!120:(a0)$) (b0) {$\phi^{b\tilde{b}}$};
        \vertex at ($(g1)!1!120:(b0)$) (c0) {$\phi^{c\tilde{c}}$};
        \diagram*{
            (g1) -- [photon] (a0),
            (g1) -- [photon] (b0),
            (g1) -- [photon] (c0),
        };
    \end{feynman}
    \end{tikzpicture}
    =& -2i y  f^{a b c} \tilde{f}^{\tilde{a} \tilde{b} \tilde{c}}.
\end{align}

For dilaton gravity, we strictly follow the convention in \cite{Plefka:2018dpa} which extensively simplifies our calculation.
Originally, the action is
\begin{align}\label{eqA:dilgrav}
    S^{\mathrm{DG}}=& -\frac{2}{\kappa^{2}} \int \dd^{4} x \sqrt{-g}\left[R-2 \partial_{\mu} \varphi \partial^{\mu} \varphi\right].
\end{align}
We will expand it in the weak field limit \eqref{eq:perturbG}.
Using the field redefinition of $\{\varphi, h_{\mu\nu}\}$ and the gauge defined in \cite{Plefka:2018dpa}, we rewrite the action as
\begin{align}\label{eqA:dilgravfinal}
    S^{\mathrm{DG}} =& \int \dd^{4} x \Big( \frac{1}{2} \partial_{\rho} h_{\mu \nu} \partial^{\rho} h^{\mu \nu} \\
    &+ \frac{\kappa}{4 \ccdot 3 !} \mathcal{V}_{123}^{\mu \alpha \gamma} \mathcal{V}_{123}^{\nu \beta \delta} h_{1 \mu \nu} h_{2 \alpha \beta} h_{3 \gamma \delta}
    + \mathcal{O}(\kappa^2, \varphi) \Big), \nonumber
\end{align}
where $\mathcal{V}_{123}^{\nu \beta \delta} = V_{123}^{\nu \beta \delta} \Big|_{k_i \to \partial_i}$ is the position space version of $V_{123}^{\nu \beta \delta}$, with the labels $1,2,3$ indicating on which $h_{\mu\nu}$ the partial derivatives should be applied.
This yields the graviton propagator
\begin{align}
    \label{eq:proph}
    \begin{tikzpicture}[baseline={(0,0)}]
    \begin{feynman}
        \vertex at (-1.85,0) (v1) {$h_{\mu \nu}$};
        \vertex at (0,0) (v2) {$h_{\rho \sigma}$};
        \diagram*{
            (v1) -- [gluon, momentum={$k$}] (v2)
        };
    \end{feynman}
    \end{tikzpicture}
     = \frac{i}{k^2} P_{\mu\nu\rho\sigma}
\end{align}
with
\begin{align}
    P_{\mu\nu\rho\sigma} = \frac{\eta_{\mu\rho} \eta_{\nu\sigma} + \eta_{\mu \sigma} \eta_{\nu \rho}}{2}
\end{align}
and the three-graviton vertex is simply
\begin{align}
    \label{eq:vertex3g}
    \begin{tikzpicture}[baseline={(current bounding box.center)}]
    \begin{feynman}
        \vertex [sdot] (g1) {};
        \vertex [above=1.2cm of g1] (a0) {$h_{\mu_1 \nu_1}$};
        \vertex at ($(g1)!1.1!120:(a0)$) (b0) {$h_{\mu_2 \nu_2}$};
        \vertex at ($(g1)!1.1!120:(b0)$) (c0) {$h_{\mu_3 \nu_3}$};
        \diagram*{
            (g1) -- [graviton] (a0),
            (g1) -- [graviton] (b0),
            (g1) -- [graviton] (c0),
            (g1) -- [opacity=0, momentum={[arrow distance=5pt, label distance=-4pt, arrow shorten=0.2]$k_1$}] (a0),
            (g1) -- [opacity=0, momentum={[arrow distance=5pt, label distance=-4pt, arrow shorten=0.2]$k_2$}] (b0),
            (g1) -- [opacity=0, momentum={[arrow distance=5pt, label distance=-4pt, arrow shorten=0.2]$k_3$}] (c0),
        };
    \end{feynman}
    \end{tikzpicture}
    \!\!\!\!\!\!\!\!= \frac{-i\kappa}{4} V_{123}^{\alpha_{1} \alpha_{2} \alpha_{3}} V_{123}^{\beta_{1} \beta_{2} \beta_{3}} \prod_{i=1}^{3} P_{\alpha_{i} \beta_{i} \mu_{i} \nu_{i}}.
\end{align}
From \eqref{eq:vertex3h} and \eqref{eq:vertex3g}, we can already see the double copy structure of the vertices.

\bibliographystyle{apsrev4-1}
\bibliography{WQCD_double_copy}
\end{document}